\documentclass{int}

\chaptertitle{Radiative Heat Transfer and Effective Transport Coefficients} 

\authors{Thomas Christen, Frank Kassubek, and Rudolf Gati}
\affiliation{ABB Schweiz AG, Corporate Research\\
Segelhofstrasse 1, CH-5405 Baden-D\"attwil}
\country{Switzerland}

\begin{document}
\maketitle

{\small
The theory of heat transfer by electromagnetic radiation is based on the radiative transfer equation (RTE)
for the radiation intensity, or equivalently on the Boltzmann transport equation (BTE) for the photon distribution.
We focus in this review article, after a brief overview on different solution methods,
on a recently introduced  approach based on truncated moment expansion. Due to the linearity of the underlying BTE,
the appropriate closure of the system of moment equations is entropy production rate minimization.
This closure provides a distribution function and the associated effective transport coefficients, like mean absorption coefficients and the Eddington
factor, for a general
number of moments. The moment approach is finally illustrated with an application of the  two-moment equations to an electrical arc.}\\[0.8cm]

\section{Introduction}
Heat radiation refers to electromagnetic radiation emitted by thermally excited degrees of freedom of matter.
If both the matter and the radiation field are at thermodynamic equilibrium, well-known relations from
thermodynamics exist between the temperature $T$, the characteristic radiation frequency $\nu$, the
energy density $E^{(eq)}$, and the pressure $p_{\rm rad}$ of the radiation field.
These are {Wien's displacement law}, $\nu = 5.88\cdot 10^{10}T$ ($\nu$ in Hz and $T$ in K),
the {caloric equation of state}, $E^{(eq)}= 7.57\cdot 10^{-16}T^4 $ (in $\rm J/m^3$), and the {thermal (or thermodynamic) equation of state},
$p_{\rm rad} = E^{(eq)}/3 $. It is then straight-forward  to derive the {Stefan-Boltzmann law}
for the power emitted by a black body, $Q= 5.67\cdot 10^{-8}T^4$ in units of W/m$^2$ (cf. \cite{LandauLifshitz}). 
In typical applications, heat radiation is relevant in the frequency range of $10^{11}-10^{16}\, {\rm Hz}$, including the upper
part of the microwave band, the infrared, the visible light, and the lower part of the ultra-violet band.\\
In many cases, be it for engineering purposes like electric arc radiation modelling, or related
fundamental scientific problems like in stellar physics, radiation is usually not at thermal equilibrium.
The present chapter of this book aims to give a focused overview on the theory
of radiative heat transfer, i.e., energy  transport by heat radiation that can be in a general nonequilibrium state,
while matter is at local thermodynamic equilibrium.
With emphasis on models based on partial differential equations for the radiation energy density, heat flux, and
(if necessary) higher order moments,
we will particularly discuss a powerful method for the determination of effective transport coefficients, which
has been recently developed by \cite{Christen2009}.
General monographs on radiative transfer are given by \cite{Chandrasekhar1960}, \cite{SiegelHowell1992}, and \cite{Modest2003}, to mention a
few.\\
In Sect. \ref{Basics} the basic definitions and equations for radiative heat transfer will be introduced.
There are two equivalent descriptions of radiation, 
either in terms of the {\em specific radiation intensity} (or {\em radiance}),
$I_{\nu}({\bf x, \bf \Omega},t)$, or the {\em photon distribution function}, $f_{\nu }({\bf x, \bf \Omega},t)$.
Here, $t$, $\bf x$, $\nu $, and $\bf \Omega$ denote time, position,
frequency, and direction (normalized wave-vector), respectively.
Frequency dependence will always be indicated by an index $\nu$.
The associated transport equations for $I_{\nu}$ and $f_{\nu}$ are named the {\em radiative transfer equation (RTE)}
and the {\em Boltzmann transport equation (BTE)}, respectively.
The number of photons in the volume element $d^3{\bf x}$ at position ${\bf x}$ and time $t$, in the frequency band
$d\nu$ at $\nu$, and in direction ${\bf \Omega}$ within the solid angle
$d{\bf \Omega }$ equals $f_{\nu }({\bf x}, {\bf \Omega},t)\, d^3{\bf x} \, d\nu \, d{\bf \Omega }$. The intensity is then given
by 
$I_{\nu }({\bf x},{\bf \Omega},t)=c h \nu f_{\nu}$, where $h\nu $ is the photon energy,
$h=\,6.626\cdot 10^{-34} Js$ is Planck's constant, and $c=2.998\cdot 10^{8}\,m/s$ is the vacuum velocity of light (cf. \cite{Tien1968}). 
$I_{\nu} $ is the energy current density per solid angle in direction ${\bf \Omega}$.\\
The RTE (or BTE) is an integro-differential equation for $I_{\nu}$ (or $f_{\nu}$) in
the 6-dimensional phase space corresponding to position, frequency,
and direction, and describes the temporal change of $I_{\nu}$ (or $f_{\nu}$) due to emission, absorption, and scattering by
the matter. 
Finding an appropriate solution is generally a highly sophisticated task, and can be significantly impeded by a complicated frequency dependence
of the radiation-matter interaction. Moreover, radiation problems in science and engineering often require a self-consistent
solution of the coupled equations for radiation and matter. For instance, a treatment of radiation in hot gases or plasma involves, besides the
RTE, the gas-dynamic balance equations for 
mass, momentum, and energy (or temperature). Despite of the recent year huge progress in computational technologies, an exact
solution of the complete set of coupled equations is still unfeasible, except for some especially simple cases. As a consequence,
in the course of time a number of methods for approximate solutions of the RTE have been developed. In Sect. \ref{Approximation}, we will therefore
briefly discuss a selected list of important approximation concepts. Methods based on
truncated momentum expansions will be emphasized, and the need of a reliable closure method for the determination of the transport
coefficients occurring in these equations will be motivated.\\
In Sect. \ref{ClosureApproaches} we will argue that a recently introduced approach for the closure based on entropy production rate is superior to other
closures used up to date. The theory of radiation in thermal equilibrium dates back to seminal work by \cite{Planck1906}. In chapter
5 of his book Planck emphasizes that photons, unlike a normal gas of massive particles,  do not interact among themselves but interaction with matter is needed
for a relaxation to the thermal equilibrium state. As is often the case in many applications of radiative transport, we will assume that the
medium, be it condensed matter, gas or plasma, is at local thermodynamic equilibrium (LTE) and can thus be described locally by
thermodynamic quantities like temperature, chemical potential, and the same. It is then the equilibration process of the photon gas to this LTE
state that determines the details of the heat transfer by radiation in the medium.
As is well-known from thermodynamics, equilibration is related to entropy production,
which plays an important role in understanding the behavior of nonequilibrium radiation (cf. \cite{Oxenius1966}
and \cite{Kroll1967}).
In fact, various authors have shown that the state of radiation is often related to optima of the entropy production rate.
Whether the optimum is a maximum or a minimum, depends on the specific details of the system under consideration,
particularly on convexity properties of the optimization problem, and particularly, the constraints.
For instance, \cite{Essex1984} has shown that the entropy production rate is minimal
in a gray atmosphere in local radiative equilibrium. Later on \cite{Essex1997} applied his approach also to neutrino radiation. 
\cite{WurfelRuppel1985} and \cite{Kabelac1994} discussed
entropy production rate maximization by introducing an effective chemical potential of the photons, related to their interaction with matter.
\cite{Santillan1998} showed that for a constraint of fixed radiation power, black bodies
maximize the entropy production rate.\\ \indent
The underlying reason for the success of entropy production rate principles has been recognized already by \cite{Kohler1948}, who has shown that
the stationary solution of the BTE that is linearized at the equilibrium distribution, satisfies a variational principle for the
entropy production rate. Kohler's principle has been widely used to determine
linear transport coefficients (cf. \cite{Ziman1956} and refs. cited in \cite{Martyushev2006}).
The important property of the RTE (or the BTE for photons)  is its linearity over the {\em whole}
nonequilibrium range, provided the interaction with the LTE-medium consists of single-photon
processes only. This linearity is thus not an approximation as it was in Kohler's work, but holds for arbitrarily large deviations from thermal
equilibrium of the photon gas. The absence of interaction between photons is thus the reason for the success of the concept beyond
small deviations of $f_{\nu}$ from equilibrium. Consequently, the entropy production rate is the appropriate basis for the determination of the
nonequilibrium distribution $I_{\nu}$ (or $f_{\nu}$) and the
effective transport coefficients for radiative heat transfer in the framework of a truncated moment expansion.
In  Sect. \ref{Results} the transports coefficients, i.e., the effective absorption constants and the Eddington factor, are calculated for some specific examples.
A practical reason for selecting moment equations for modelling radiative transfer is the convenience of having a set of structurally similar equations for
the simulation of the complete radiation-hydrodynamics problem. Both the hydrodynamic equations for matter and the moment equations for radiation
are hyperbolic partial differential equations and can thus be solved on the same footing. In Sect. \ref{Hyperbolicity} gives some remarks on the requirement of hyperbolicity.
For numerical simulations boundary conditions must be specified, these will be discussed in Sect. \ref{BoundaryConditions}.
Finally, Sect. \ref{Arc} will then provide
some simulation results for a simplified example of electric arc radiation.

\section{Basics of Radiative Heat Transfer in Matter}
\label{Basics}
The radiation intensity, $I_{\nu }({\bf \Omega})$, is governed
by the radiative transfer equation (RTE),
\begin{equation}
\frac{1}{c}\partial _{t} I_{\nu } + {\bf \Omega \cdot { \nabla}} I_{\nu } =  \kappa {_\nu} (B_{\nu} - I_{\nu})
+ \sigma {_\nu}\left( \frac{1 }{4\pi}\int _{S^{2}} d \tilde \Omega \, p_{\nu}(\Omega , \tilde \Omega)\,I_{\nu} (\tilde \Omega)-   I_{\nu}\right)
\;\;,
\label{RTE}
\end{equation}
which has to be solved in a spatial region defined by the physical problem under consideration.
Phase coherence and interference effects are disregarded when considering thermal radiation, and
we will also not consider polarization effects.
The BTE is simply obtained by a replacement of $I_{\nu}$ by $f_{\nu }$ in the RTE.
The left hand side gives the total rate of change of $I_{\nu}({\bf \Omega})$, divided by $c$, along the propagation direction ${\bf \Omega}$.
This change must be equal to the expression on the right hand side,
which consists of a sum of specific source and sink terms due to the radiation-matter interaction. In the absence of any interaction, e.g., in
vacuum, the right hand side vanishes, which describes the so-called (free) streaming limit associated with a radiation beam, or
the ballistic propagation of the photons.
In the presence of interaction, however, photons are generated by emission and annihilated by absorption,
described by $ \kappa {_\nu} B_{\nu}$ and $- \kappa {_\nu} I_{\nu}$, respectively. Here, $B_{\nu}$
is the Planck function for thermal equilibrium,
\begin{equation}
B_{\nu }=\frac{2h\nu ^{3}}{c^{2}} n^{(eq)} \;\; ,
\label{Planckdistribution}
\end{equation}
where
\begin{equation}
n^{(eq)}_{\nu} =\frac{1}{\exp (h\nu /k_{B}T)-1} \;\; 
\label{BoseEinstein}
\end{equation}
is the Bose-Einstein distribution for thermal equilibrium photons (cf. \cite{LandauLifshitz})
with the Boltzmann constant $k_{B}=1.381\cdot 10^{-23}J/K$ and the local temperature $T=T(\bf x)$ of the LTE medium. The coefficient 
$\kappa _{\nu}$ is the macroscopic spectral absorption coefficient in units of $1/m$, and is generally a sum of products of particle
densities, absorption cross-sections, factors $[1-\exp(-h\nu/k_{B}T)]$,
and depends thus not only on frequency but also on the partial pressures of the present species and the temperature.
Often, opacities referring to $\kappa _{\nu}/\rho$ are discussed in the literature, where $\rho$ is the mass density of the matter. 
The macroscopic $\kappa _{\nu}$ includes
spontaneous as well as induced emission (cf. \cite{Tien1968}).
Additionally to inelastic absorption-emission processes, Eq. (\ref{RTE}) includes elastic (or so-called coherent or conservative) scattering.
Incoming photons of frequency $\nu$ from all directions $\tilde \Omega$ are
scattered with probability $p_{\nu}(\bf \Omega , {\bf \tilde \Omega})$ into direction $\bf \Omega $. It is assumed that the
phase-function $p_{\nu} (\Omega , \tilde \Omega) $ obeys symmetry relations associated with reciprocity, depends
only on the cosine between the directions ${\bf \tilde \Omega}$ and ${\bf \Omega}$ (cf. \cite{Chandrasekhar1960}), and 
is normalized, $(4\pi)^{-1}\int _{S^2}\,d\tilde \Omega \, p_{\nu}(\Omega , \tilde \Omega) =1$. Here, the
$\Omega $-Integration extends over $S^2$, which denotes the unit sphere associated with the full solid angle.
The strength of the scattering process is quantified by the spectral scattering coefficient $\sigma _{\nu}$ in units of $1/m$. The ratio
$\sigma _{\nu} /(\kappa _{\nu} + \sigma_{\nu})$ gives the probability that a collision event is a scattering process, and is sometimes called
the {\em (single-scattering) albedo}. The mean free path of the photons is the inverse of $\kappa _{\nu} + \sigma_{\nu}$.\\
Because the bracket proportional to $\sigma {_\nu}$ in Eq. (\ref{RTE}) vanishes for ${\bf \Omega}$-independent $I_{\nu}$,
the RTE can be written in the simple form
\begin{equation}
\frac{1}{c}\partial _{t} I_{\nu } + {\bf \Omega \cdot { \nabla}} I_{\nu } =  \mathcal{L}(B_{\nu}-I_{\nu}) 
\;\;,
\label{RTEshort}
\end{equation}
where the linear, self-adjoint, positive semi-definite\footnote{Note that a negative
eigenvalue would immediately lead to an instability.} operator $\mathcal{L}$ is defined by the right hand side of
Eq. (\ref{RTE}) and consists of an algebraic term and an integral term.\\

The RTE has to be solved with appropriate initial conditions, $I_{\nu}({\bf x},{\bf \Omega}, t=0)$, and boundary conditions
on the surface of the spatial domain under consideration. Because the RTE is a first order differential equation, the determination of
each ray requires the knowledge of $I_{\nu}({\bf \Omega})$ on the domain surface and in directions ${\bf \Omega}$ pointing into the domain.
The behavior of the boundary is characterized by the radiation it emits, and the way it reflects impinging radiation. 
If one denotes the emittance of the boundary at position ${\bf x_{w}}$ by $\epsilon ({\bf x_{w}})$, 
the reflectivity by $r ({\bf x_{w}})$, and the normal vector of the boundary surface by ${\bf n ({\bf x_{w}})}$,
the boundary condition generally reads (cf. \cite{Modest2003})
\begin{equation}
I_{\nu}({\bf x_{w}},{\bf \Omega},t)= \epsilon ({\bf x_{w}})  B_{\nu}({\bf x_{w}}) + \int _{{\bf n ({\bf x_{w}})}\cdot {\bf \tilde \Omega} \leq 0}
d {\bf \tilde \Omega} 
\mid {\bf n ({\bf x_{w}})}\cdot {\bf \tilde \Omega} \mid r ({\bf x_{w}},{\bf \Omega} ,{\bf \tilde \Omega} ) I_{\nu}({\bf x_{w}},{\bf \tilde \Omega},t )\,  \;\; .
\label{boundaryconditionRTE}
\end{equation}
The integration runs over all ${\bf \tilde \Omega }$ associated with radiation coming from the bulk domain towards the surface,
while ${\bf \Omega }$ is  pointing into the domain. For a smooth surface
where a normal vector ${\bf n ({\bf x_{w}})}$ can be defined, this solid angle corresponds to half of the sphere $S^{2}$.
This general boundary condition can be simplified for special limit cases. For instance, a black surface has $r =0$ and $\epsilon =1$,
a diffusively reflecting surface has $r ({\bf x_{w}},{\bf \Omega} ,{\bf \tilde \Omega} )= r ({\bf x_{w}})/\pi$, and a specularly
reflecting surface has $r ({\bf x_{w}},{\bf \Omega} ,{\bf \tilde \Omega} )\propto \delta ({\bf \Omega _{s}} - {\bf \tilde \Omega}) $, where
${\bf \Omega_{s}} = {\bf \Omega} -2({\bf \Omega } \cdot {\bf n} ){\bf n} $ is the direction from which the ray must hit the surface in order to travel into the direction of
${\bf \Omega } $ after specular reflection.\\

We conclude this section by listing the basic equations for the LTE matter to which radiation is coupled.
In general, LTE implies that at each point in space,
the caloric and thermodynamic equations of state are locally valid. The respective equations relate the specific energy $e=e(\rho, T)$ and the
pressure $p=p(\rho ,T )$ to the mass density $\rho $ and the temperature $T$ of the matter. The spatio-temporal dynamics
of the thermodynamic variables and, if relevant, the flow velocity ${\bf u}$, is then given by the hydrodynamic
balance equations for mass, momentum, and energy. For a single component (non-relativistic) medium
\begin{eqnarray}
 \partial _{t} \rho  + {\bf \nabla} \cdot (\rho {\bf u})  & = & \dot \rho \label{massbalance} \\ 
  \partial _{t} (\rho {\bf u}) + {\bf \nabla} \cdot {\bf \Pi_{\rm mat}} &= & {\bf f} \label{momentumbalance} \\
 \partial _{t} (\rho e_{\rm tot}) + {\bf \nabla} \cdot {\bf j_{e}}  &= & W \label{heatbalance} 
\end{eqnarray}
where ${\bf \Pi_{\rm mat}}$, ${\bf j_{e}}$, and $e_{\rm tot}=e+{\bf u}^{2}/2$ are the momentum stress tensor,
the energy flow density, and the total energy density.
Together with the equations of state,
Eqs. (\ref{massbalance})-(\ref{heatbalance}) constitute seven equations for the seven
variables $\rho$, $p$, $T$, $e$, and ${\bf u}$. The right hand sides, $\dot \rho$,
${\bf f}$, and $W$ are the
mass source density, the force density, and the heat power density, respectively.
The effect of radiation on matter may occur in these three source terms. For instance, a mass source may appear at a solid wall due to ablation by radiation (see, e.g.
\cite{Christen2007}), and the radiation pressure may act as a force (cf. \cite{Mihalas1984}). These two effects are often
negligible in engineering applications or are important only in
special cases, like ablation arcs as discussed by \cite{Nordborg2008}. However,
the heat exchange described by $W$ can in general not be disregarded, and will play an important role in the theory
below. The back-coupling of the matter on radiation, as mentioned before, occurs in the expressions on the right hand side of
Eq. (\ref{RTE}), which depend generally on $\rho$ (or $p$) and $T$. An extensive monograph on radiation hydrodynamics is provided by
\cite{Mihalas1984}, and a short introduction that fits well to the present chapter is given by the lecture notes of \cite{Pomraning1982}.


\section{Approximation Methods}
\label{Approximation}
The extreme difficulties to solve the RTE exactly for real systems caused the development of various approximation
methods. There are two additional reasons for the use of approximations. First, in many cases the behavior of the matter is of interest, while
it is sufficient to consider the radiation as a means of (nonlocal) interaction; hence only the radiative heat flux is needed,
which enters the power balance equation for the matter via the heat power density $W$. As $W$
equals the negative divergence of the radiation energy flux density, a radiation model would be convenient that is
confined to this flux and to the lower order moments, which is here a single one, namely the radiation energy density.
Secondly, radiation often behaves in two different specific ways. In a transparent medium absorption and scattering are weak,
and radiation propagates as beams; full absence of interaction with matter refers to the so-called {\em free streaming limit}.
In an opaque medium, on the other hand, absorption, emission and/or scattering is strong, and the radiation 
diffuses isotropically. In the extreme diffusive limit the {\em Rosseland  diffusion approximation} applies, where radiative transfer
is modelled by an effective heat conductivity of the matter (cf. \cite{SiegelHowell1992}) given by $16\sigma_{SB}T^{3}/3\sigma _{F}^{\rm (eff)}$. Here
$\sigma _{SB}=2\pi^5k_{B}^{4}/15h^{3}c^{2}=5.67\cdot 10^{-8}\,W/m^2K^4$ is the Stefan-Boltzmann constant and $\sigma _{F}^{\rm (eff)}$
is the Rosseland mean absorption to be discussed later. 
For the two behaviors a ballistic (beam-like) and a diffusive description, respectively, are the appropriate
'zero order' models with effective transport coefficients, and deviations from the limits may be treated by small corrections. Models based on
one of these two limit cases can strongly reduce the computational effort. However, in many real systems radiative transfer is
in between these limits such that more sophisticated methods must be involved.\\
In the following, a short list of some relevant approximation methods is given. The selection is not complete,
as other approaches exist, like ray tracing and radiosity-irradiosity methods (\cite{Rey2006}), or some
rather heuristic methods like the $P_{1/3}$-approximation discussed by \cite{Olson2000} and \cite{Simmons2000}.
Furthermore, we will not discuss the issue of discretization methods concerning position space like finite differences, volumes, or elements;
although this field would require special recognition (cf. \cite{Arridge2000} and Refs. cited therein) it is beyond the purpose of
this chapter. Needless to say that there is not a
unique best method but every approach has its advantages and disadvantages for practical use, and the appropriate choice depends usually on the
problem under consideration. Exhaustive overviews can be found, e.g.,
in \cite{DuderstadtMartin1979}, \cite{SiegelHowell1992} and literature cited in the
following three subsections. Subsequently, we will then focus in subsection \ref{Moment}
on approximations based on moment expansions, and particularly on the closure of the moment
equations that will be discussed in Sect. \ref{ClosureApproaches}.
%
\subsection{Net Emission}
\label{NEM}
The {\em net emission approximation} is probably the most simplistic radiation model. It assumes a semi-empirical function
$W(T,p,{\bm \zeta})$ in Eq. (\ref{heatbalance}). Additionally to temperature and pressure, it depends on parameters ${\bm \zeta}$
of the radiating object. It is sometimes used, for instance, in computational fluid
dynamics simulations of electrical arcs (cf. \cite{Lowke1970},
\cite{Zhang1987}, \cite{Seeger2006}), where the only parameter ${\zeta}$ is the arc radius. Although such a description is very convenient in numerical simulations
and sometimes even provides useful results, it is obviously oversimplifying and without any rigor. Furthermore, reliable accuracy requires, for the determination of
the function $W(T,p,{\zeta})$, a parameter study based on a more fundamental radiation model or on elaborate experiments.
%

\subsection{Monte Carlo}
\label{MonteCarlo}
{\em Monte Carlo simulations} refer to random sampling methods (see, for instance \cite{Yang1995} and \cite{DuderstadtMartin1979}),
which are based on computer simulations of a number of photons. Their deterministic dynamics corresponds to the ballistic
motion with speed of light. Emission, absorption, and scattering processes are simulated in a probabilistic way by appropriately determined random
numbers for the various processes. Those include, of course, the interaction with boundaries of the spatial domain. Final results, like the
radiation intensity, are determined by averages over many particles. The Monte Carlo concept is rather simple, which leads to
a number of advantages of this method, as discussed by \cite{Yang1995}. Efficient applications make use of specifically improved
schemes like implicit Monte Carlo or special versions thereof
(cf. \cite{Brooks1986} and \cite{Brooks2005}).

\subsection{Discrete Ordinates}
\label{DOM}
The {\em discrete ordinates method} ({\em DOM}) considers a finite number of rays passing
at every (discrete) space point. If a number $N_{D}$ of
direction vectors ${\bf \Omega }_{k}$, $k=1,...,N_{D}$ is selected, one has $I_{\nu}=\Sigma _{k}^{N_D}I_{\nu}^{(k)}\delta ({\bf \Omega }-{\bf \Omega }_{k})$,
such that a set of $N_{D}$ partially coupled RTE-equations for the different directions and frequencies must be solved. 
The right hand side of these equations, say $\mathcal{\tilde L}(B_{\nu}-I_{\nu}^{(k)})$, contains not an integral as Eq. (\ref{RTE}) but a weighted sum.
As reasonable minimum values for $ N_{D}$ in 3-dimensional realistic geometries are of the order of 10, the computational effort is still
large. For too small $N_{D}$ an artifact called "ray effect" may occur, referring to
spatial oscillations in the energy density. Another error known as "false scattering" or "false diffusion", is due to the discretization of position space 
and is linked in a certain way to the ray effect as discussed in \cite{Rey2006}.\\
Some further developments based on DOM exist, which make use of a decomposition and discretization of the angular space into a finite set of directions,
i.e. a finite partition of the unit sphere $S^2$. The methods of {\em partial characteristics} (\cite{Aubrecht1994})
and of {\em partial moments} (\cite{Frank2006}) are examples, the latter being mentioned again in the next section.
Last but not least, we mention that is has been proven that the DOM is equivalent, under certain conditions,
to the P-N method (cf. \cite{BarichelloSiewer1998} and \cite{Cullen2001}), which is a special kind of the moment approximations to be
discussed in the next subsection. 
%
%
%
\subsection{Moment Expansions}
\label{Moment}
Radiation modelling in terms of moments of the distribution $I_{\nu}$ (or $f_{\nu}$) is convenient because the radiation
is coupled to the LTE matter in Eqs. (\ref{massbalance})-(\ref{heatbalance}) via the first three (angular) moments. Moment expansions
can be formulated in a rather general manner (cf. \cite{Levermore1996} and \cite{Struchtrup1998}). In the following,
we define moments based on $I_{\nu}$ by\footnote{We mention that a moment corresponding to the photon number (obtained by integration over $f_{\nu}$)
does not appear, partly because the photon number is not a conserved quantity.}
\begin{eqnarray}
E & = & \int _{0}^{\infty}  d\nu \, E_{\nu} = \int _{0}^{\infty} d\nu \frac{1}{c}\int _{S^2}  d\Omega \, I_{\nu} \;\; ,\label{Edef} \\
{\bf F} & = & \int  _{0}^{\infty}  d\nu \,  {\bf F}_{\nu}=\int _{0}^{\infty} d\nu  \frac{1}{c}\int _{S^2} d\Omega \, {\bf \Omega} \,  I_{\nu}\;\; , \label{Fdef} \\
{\bf \Pi} & = & \int  _{0}^{\infty}  d\nu \, {\bf \Pi }_{\nu} = \int _{0}^{\infty} d\nu  \frac{1}{c}\int _{S^2} d\Omega \,\, {\bf \Omega: \Omega}\, I_{\nu} \;\; ,\label{Pdef} \\
... & = & ... \;\;\; \;\; , \nonumber  
\end{eqnarray}
with $({\bf \Omega: \Omega})_{kl} = {\Omega _{k} \Omega _{l}}$.
The last line indicates that an infinite number of moments exist in general.
$E_{\nu}$, ${\bf F}_{\nu}$, and ${\bf \Pi}_{\nu}$ are, respectively, the monochromatic energy density, radiative flux, and stress or pressure tensor
of the radiation. For convenience, the prefactor ($c^{-1}$) is chosen in all definitions such that the moments have the same
units of a spectral energy density.  
Similarly, $E$, ${\bf F}$, and ${\bf \Pi}$ are the spectrally integrated energy density,
radiative flux, and pressure tensor. In the present units $F$ has the meaning of energy density associated with the average directed motion of the photons, and $E$ of
the total energy density composed of directed and thermal fluctuation parts. Hence, $F=\mid {\bf F}\mid \leq E$, which will be important below.\\
In thermal equilibrium all fluxes vanish. Then ${\bf F}_{\nu}=0$, the stress tensor is proportional to the unit tensor with diagonal elements  
$E^{(eq)}/3$, and the energy density is given by
\begin{equation}
E^{(eq)} = \int _{0}^{\infty} 4\pi \, d\nu\, B_{\nu }=\frac{4\sigma_{SB}}{c}T^{4} \;\; .
\label{SB1}
\end{equation}
The purpose of a moment expansion is to derive from the RTE or BTE balance equations for the moments,
either for each frequency $\nu $, or for groups of frequencies or frequency bands, or for the full, integrated spectral range.
Multiplication of the RTE with products and/or powers of $\Omega _{k}$'s, and
integration over the solid angle gives for the moments $E_{\nu}$, ${\bf F}_{\nu}$, etc.
\begin{eqnarray}
\frac{1}{c}\partial _{t}E_{\nu} + {\bf \nabla \cdot F}_{\nu}  & = &  \frac{1}{c}\int _{S^{2}}d\Omega \,\mathcal{L}(B_{\nu}-I_{\nu})  \;\; , \label{Enueq} \\
\frac{1}{c} \partial _{t}{\bf F}_{\nu}+ {\bf \nabla \cdot \Pi}_{\nu} & = & \frac{1}{c} \int _{S^{2}}d\Omega \,{\bf \Omega}\, \mathcal{L}(B_{\nu}-I_{\nu})  \;\; , \label{Fnueq} 
\end{eqnarray}
etc., where only the first two equations are listed for convenience, but the list still contains an infinite number for all moments and for all frequencies.
Practical usability calls then for a two-fold approximation. First, the list of moments, and thus moment equations should be truncated by considering
only the $N$ first moment equations. Secondly, the frequency space should be discretized or partitioned in some way, in order to end up with a finite set.  
If the spectrum allows a division into a number of well defined frequency bands with approximately constant $\kappa _{\nu}$ and $\sigma _{\nu}$,
or a grouping of different frequencies together according to similar values of $\kappa _{\nu}$ and $\sigma _{\nu}$,
one can average the equations over such partitions. The associated methods are sometimes named multi-group, multi-band, or multi-bin
methods.  For details, we refer the reader to \cite{Turpault2005}, \cite{Ripoll2008}, \cite{Nordborg2008}, and the literature cited therein.
In the following we will consider the equations for the spectrally averaged quantities, which are obtained by
integration of Eqs. (\ref{Enueq}), (\ref{Fnueq}), etc., over frequency
\begin{eqnarray}
\frac{1}{c}\partial _{t}E + {\bf \nabla \cdot F}  & = & P_{E}=  \frac{1}{c} \int _{0}^{\infty} d\nu\int _{S^{2}}d\Omega \,\mathcal{L}(B_{\nu}-I_{\nu})   \;\; , \label{Etdef} \\
\frac{1}{c} \partial _{t}{\bf F}+ {\bf \nabla \cdot \Pi} & = & {\bf P_{F}} =  \frac{1}{c} \int _{0}^{\infty} d\nu \int _{S^{2}}d\Omega  \;\; , \,{\bf \Omega}\,
\mathcal{L}(B_{\nu}-I_{\nu}) \label{Ftdef}
\end{eqnarray}
etc., where the right hand sides define $P_{E} $ and ${\bf P_{F}}$, etc.
These quantities are still functionals of the unknown function $I_{\nu}$. All moments, on the other hand,
are variables that are determined by the full (still infinite) set of partial differential equations, provided reasonable initial and boundary conditions
are given.\\
Now we perform a truncation by using only the first $N$ moment equations. The first $N$ moments would then be determined by the solution of these equations, if
the right hand sides ($P_{E} $, ${\bf P_{F}}$, etc) and the $N+1$'th moment were known.
In the following section we will discuss closure methods that determine these unknowns that are supposed to be functions of the
$N$ moments. Prior, however, we remark that
instead of using products of Cartesian coordinates of ${\bf \Omega }$, one may equivalently consider a representation in terms of
spherical coordinates $(\theta,\phi)$. The radiation density is then expanded in spherical harmonics $Y^{m}_{l}(\theta,\phi)$. If truncated,
this approximation corresponds to the P-N approximation (cf. \cite{SiegelHowell1992}).
The prominent P-1 approximation (cf. \cite{SiegelHowell1992}), for instance,
refers to a truncation of the Eqs. (\ref{Enueq}) and (\ref{Fnueq}) (or Eqs. (\ref{Etdef}) and (\ref{Ftdef}))
after the second equation and considers an isotropic $\Pi _{\nu}$ (or $\Pi$) with
diagonal elements equal to  $E_{\nu }/3$ (or $E/3$).\\  
We also mention again the partial moment approximation (cf. \cite{Frank2006}),
where the approaches of DOM and moment expansion are combined in a smart way. As the DOM discretizes the angular space
in different directions, the partial moment method selects partitions $\mathcal{A}$ of the unit sphere
$S^{2}$ and defines partial moments $E_{\nu}^{(\mathcal{A})}$, ${\bf F}_{\nu}^{(\mathcal{A})}$, ${\bf \Pi}_{\nu}^{(\mathcal{A})}$, etc,
where the solid angle integration is performed only over $\mathcal{A}$ instead of the whole $S^{2}$. The most simple but nontrivial
partial moment model refers to the forward and backward traveling waves in a one-dimensional position space, where the integration occurs
over the two half-spheres associated with forward and backward directions. According to \cite{Frank2007},
this method is able to resolve a shock-wave artifact occurring for
counter-propagating and interpenetrating radiation beams.

\section{Closure Approaches}
\label{ClosureApproaches}
The quality of the moment approximation depends on the number of moments taken into account, and on the specific closure concept.
A closure of a truncated moment expansion requires in principle knowledge of $I_{\nu}$.
A simplification occurs if $\kappa _{\nu}$, $\sigma _{\nu}$, and $p_{\nu}$ are assumed to be constant (gray matter).
The right hand sides of the moment equations strongly simplify as
they can be directly expressed in terms of these constants and linear expressions of the moments.
But in general matter is non-gray, and the absorption and
scattering spectra can be extremely complex. Furthermore, the $N+1$'th
moment remains still unknown even for gray matter.  In the sequel we will discuss a few practically relevant
closure methods. We will then argue that the preferred closure
is given by an entropy production principle.\\
For clarity we will consider the two-moment example; generalization to an arbitrary number of
moments is straight-forward.
The appropriate number of moments is influenced by the geometry and the optical density of the matter. For symmetric geometries,
like plane, cylindrical, or spherical symmetry, less moments are needed than for complex arrangements with shadowing corners, slits
and the same. For optically dense matter, the photons behave diffusive, which can be modelled well by a low number of moments, as will be
discussed below. For transparent media, beams, or even several beams that might cross and interpenetrate, may occur, which
makes higher order or multipole moments necessary.

\subsection{Two-Moment Example}
\label{TwoMoment}
The unknowns are $P_{E} $, ${\bf P_{F}}$, and ${\bf \Pi}$, which may be functions of the two moments
$E$ and ${\bf F}$.  For convenience, we will write
\begin{eqnarray}
P_{E} & = &\kappa _{\rm E}^{\rm (eff)} (E^{(eq)}-E)  \;\; , \label{kappaEeff} \\ 
{\bf P_{F}}  & = & -\kappa _{\rm F}^{\rm (eff)}{\bf F}  \;\; . \label{kappaFeffs}
\end{eqnarray}
where we introduced the effective absorption coefficients $\kappa _{\rm E} ^{\rm (eff)}$ and $\kappa _{\rm F} ^{\rm (eff)}$ that are generally
functions of $E$ and ${\bf F}$. 
Because the second rank tensor ${\bf \Pi }$ depends only on the scalar $E$ and the vector ${\bf F}$, by symmetry reason
it can be written in the form
\begin{equation}
\Pi _{nm} = E\left( \frac{1-\chi}{2}+ \frac{3\chi-1}{2}\,\frac{F_{n}F_{m}}{F^2} \right)\;\; ,
\label{Eddingtondel}
\end{equation}
where the {\em variable Eddington factor (VEF)} $\chi $ is a function of $E$ and ${\bf F}$. Assuming that the underlying space is
isotropic, $\kappa _{E}^{\rm (eff)}$, $\kappa _{F}^{\rm (eff)}$, and $\chi$ can be expressed as functions of $E$ and
\begin{equation}
v =  \frac{F}{E}\;\; ,
\label{vdef}
\end{equation}
with $F=\mid {\bf F}\mid$. Obviously it holds $0\leq v \leq 1$, with $v=1$ corresponding to
a fully directed radiation beam (free streaming limit). According to \cite{Pomraning1982}, the additional $E$ dependence of suggested or derived
VEFs often appears via an effective $E$-dependent single scattering albedo, which equals, e.g. for gray matter,
$(\kappa E^{(eq)}+\sigma E)/(\kappa +\sigma)E $.\\
The task of a closure is thus to determine the {\em effective transport coefficients},
i.e., {\em effective mean absorption coefficients} $\kappa _{E}^{\rm (eff)}$, $\kappa _{F}^{\rm (eff)}$, and the VEF $\chi$ of $E$ and $F$ (or $v$).
This task is of high relevance in various scientific fields, from terrestrial atmosphere physics and astrophysics to engineering plasma
physics. 

%
\subsection{Exact Limits and Interpolations}
\label{Interpolations}
In limit cases of strongly opaque and strongly transparent matter, analytical expressions for the effective absorption coefficients are often used,
which can be determined in principle from basic gas properties (see, e.g., \cite{Aburomia1967} and \cite{Fuss2002}).
In an optically dense medium radiation behaves diffusive and isotropic, and is near equilibrium with respect to LTE-matter. 
The effective absorption coefficients are given by the
so-called {\em Rosseland average} or {\em Rosseland mean} (cf. \cite{SiegelHowell1992})
\begin{equation}
\kappa _{E}^{\rm (eff)} = \langle\kappa _{\nu} \rangle_{\rm Ro} := \frac{\int_{0}^{\infty} \,d\nu \;\nu^{4}\partial _{\nu } n^{(eq)}_{\nu}}
{\int_{0}^{\infty} \,d\nu \, \nu ^{4} \kappa _{\nu}^{-1} \partial _{\nu } n^{(eq)}_{\nu}}  \;\; ,
\label{RosselandmeanE}
\end{equation}
where $\partial _{\nu }$ denotes differentiation with respect to frequency, and 
\begin{equation}
\kappa _{F}^{\rm (eff)} =\langle\kappa _{\nu}+\sigma _{\nu}\rangle_{\rm Ro} \;\; .
\label{RosselandmeanF}
\end{equation}
The Rosseland mean is an average of inverse rates, i.e., of times, and must thus be associated with consecutive processes.
A hand-waving explanation is based on the strong mixing between different frequency modes by the many absorption-emission processes in the
optically dense medium, due to the short photon mean free path.\\
Isotropy of ${\bf \Pi}$ implies for the Eddington factor $\chi = 1/3$. Indeed, because $\sum \Pi_{kk} = E$, one has
then ${\bf \Pi}_{kl}=\delta _{kl}E/3$,
where $\delta _{kl}$ ($=0$ if $k\neq l$ and $\delta _{kl}=1$ if $k=l$) is the Kronecker delta. 
With these stipulations, Eqs. (\ref{Etdef}) and (\ref{Ftdef}) are completely defined and can be solved.\\
In a strongly scattering medium ($\sigma _{\nu} \gg \kappa _{\nu} $), where $\bf F$ relaxes quickly to its quasi-steady state, one may further assume
 ${\bf F} = -\nabla E / 3\kappa _{F}^{\rm (eff)}$ for appropriate time scales.  Hence Eq. (\ref{Etdef}) becomes 
\begin{equation}
\frac{1}{c}\partial _{t}E -   \nabla \cdot \left( \frac{ \nabla  E } {3\kappa _{F}^{\rm (eff)}} \right) = 
\kappa _{E}^{\rm (eff)}(E^{(eq)}-E) \;\;,
\label{diffusion1}
\end{equation}
which has the form of a reaction-diffusion equation. For engineering applications, $E$ often relaxes much faster than all other hydrodynamic
modes of the matter, such that the time derivative of Eq. (\ref{diffusion1}) can be disregarded by assuming full
quasi-steady state of the radiation.
Equation (\ref{diffusion1}) is then equivalent to an effective steady state gray-gas P-1 approximation.\\

For transparent media, in which the radiation beam interacts weakly with the matter,
the  Planck average is often used,
\begin{equation}
\langle \kappa _{\nu} \rangle_{\rm Pl} = \frac{\int_{0}^{\infty}
\, d\nu \,\nu^{3} \kappa _{\nu} n_{\nu}^{(eq)} }{\int_{0}^{\infty} \, d\nu \, \nu^{3} n_{\nu}^{(eq)}} \;\;.
\label{Planckmean}
\end{equation}
In contrast to the Rosseland mean, the Planck mean averages the rates and can thus be associated with parallel processes, because scattering is
weak and there is low mixing between different frequency modes.
In contrast to the Rosseland average, the Planck average is dominated by the largest values of the rates.
Although in this case radiation is generally not isotropic, there are special cases where
an isotropic ${\bf \Pi}$ can be justified; an example discussed below is the $v\to 0$ limit in the emission limit $E/E^{(eq)}\to 0$.
But note that $\chi = 1$ often occurs in transparent media, and consideration of the VEF is necessary.\\

In the general case of intermediate situations between opaque and transparent media,
heuristic interpolations between fully diffusive and beam radiation are sometimes performed.
Effective absorption coefficients have been constructed heuristically by \cite{Patch1967}, or
by \cite{Sampson1965} by interpolating Rosseland and Planck averages.\\
The consideration of the correct stress tensor is even more relevant, because the simple $\chi = 1/3$
assumption can lead to the physical inconsistency $v>1$. A common method to solve this problem 
is the introduction of flux limiters in diffusion approximations,
where the effective diffusion constant is assumed to be state-dependent
(cf. \cite{Levermore1981}, \cite{Pomraning1981}, and \cite{Levermore1984}, and
Refs. cited therein). A similar approach in the two-moment model is the use of a heuristically constructed VEF.
A simple class of flux-limiting VEFs is given by 
\begin{equation}
\chi = \frac{1+2v^j}{3} \;\; ,
\label{VEFsimple}
\end{equation}
with positive $j$. These VEFs depend only on $v$, but not additionally separately on $E$.
The cases $j=1$ and $j=2$ are attributed to \cite{Auer1984} and \cite{Kershaw1976}, respectively.
While the former strongly simplifies the moment equations by making them piecewise linear,
the latter fits quite well to realistic Eddington factors, particularly for gray matter.
%

\subsection{Maximum Entropy Closure}
\label{MaxEP}
An often used closure is based on {\em entropy maximization} (cf. \cite{Minerbo1978}, \cite{Anile1991},
\cite{Cernohorsky1994}, and \cite{Ripoll2001}).\footnote{In part of the more mathematically
oriented literature, the entropy is defined with different sign and the principle is called "minimum entropy closure".}
This closure considers the local radiation entropy as a functional of $I_{\nu}$. The entropy is defined at each position ${\bf x}$ and
is given by (cf. \cite{LandauLifshitz}, \cite{Oxenius1966}, and \cite{Kroll1967})
\begin{equation}
S_{\rm rad} [I_{\nu}]  = -k_{B}  \int d\Omega\,d\nu \frac{2\nu ^2}{c^3}\left( n_{\nu} \ln n_{\nu}-(1+n_{\nu})\ln(1+n_{\nu}) \right),
\label{radiationentropy}
\end{equation}
where 
\begin{equation}
n_{\nu}({\bf x},{\bf \Omega})=\frac{c^2I_{\nu}}{2h\nu^3}
\label{photonnumber}
\end{equation}
is the photon distribution for the state ($\nu, {\bf \Omega}$).\footnote{Note the
simplified notation of a single integral symbol $\int $ in Eq. (\ref{radiationentropy}) and in the following, which is to be associated
with full frequency and angular space.} At equilibrium (\ref{photonnumber}) is given by (\ref{BoseEinstein}).
$I_{\nu}$ is then determined by maximizing $ S_{\rm rad} [I_{\nu}] $,
subject to the constraints of fixed moments given by Eqs. (\ref{Edef}), (\ref{Fdef}) etc.
This provides $I_{\nu}$ as a function of $\nu$, $\Omega$,  $E$ and ${\bf F}$. If restricted to 
the two-moment approximation, the approach is sometimes called the M-1 closure. It is generally applicable
to multigroup or multiband models (\cite{Cullen1980}, \cite{Ripoll2004}, \cite{Turpault2005}, \cite{Ripoll2008}) and partial moments (\cite{Frank2006},
\cite{Frank2007}), as well as for an arbitrarily large number of (generalized) moments (\cite{Struchtrup1998}).
It is clear that this closure can equally be applied to particles obeying Fermi statistics (see
\cite{Cernohorsky1994} and \cite{Anile2000}).\\
Advantages of the maximum entropy closure are the mathematical simplicity and
the mitigation of fundamental physical inconsistencies (\cite{Levermore1996} and \cite{Frank2007}). In particular,
there is a natural flux limitation by yielding a VEF with correct limit behavior in both
isotropic radiation ($\chi \to 1/3$) and free streaming limit ($\chi \to 1$):
\begin{equation}
\chi _{\rm ME} = \frac{5}{3}-\frac{4}{3} \sqrt{1-\frac{3}{4} v^2} \;\; 
\label{Levermore}
\end{equation}
that depends only on $v$.
Furthermore, because the optimization problem is convex\footnote{Convexity refers here to the mathematical entropy
definition with a sign different from Eq. (\ref{radiationentropy})}, the uniqueness of the solution is ensured and, as shown
by \cite{Levermore1996}, the moment equations are hyperbolic, which is important because otherwise the radiation model
would be physically meaningless.
The main disadvantage is that the maximum entropy closure is unable to
give the correct Rosseland mean in the near-equilibrium limit, and can thus not be correct.
For example, for $\sigma _{\nu}\equiv 0$ the near-equilibrium effective absorption coefficients
are given by (\cite{Struchtrup1996})
\begin{equation}
\langle \kappa _{\nu} \rangle_{\rm ME} = \frac{\int_{0}^{\infty}
\, d\nu \,\nu ^{4} \kappa _{\nu}\partial _{\nu } n^{(eq)}_{\nu} }{\int_{0}^{\infty} \, d\nu \, \nu ^{4} \partial _{\nu } n^{(eq)}_{\nu}} \;\;,
\label{MEmean}
\end{equation}
which is a Planck-like mean that averages
$\kappa _{\nu}$ instead of averaging its inverse. It is only seemingly surprising that the maximum entropy closure
is wrong even close to equilibrium. This closure concept must fail in general, as \cite{Kohler1948} has proven that for the linearized
BTE the {\em entropy  production rate}, rather than the {\em entropy}, is the quantity that must be optimized.
Both approaches lead of course to the correct equilibrium distribution. But the quantity responsible for transport is the
first order deviation $\delta I_{\nu}=I_{\nu} - B_{\nu}$, which is determined by the entropy production and not by the entropy.
Moreover, it is obvious that Eq. (\ref{radiationentropy}) is explicitly independent of the radiation-matter interaction.
Consequently, the distribution resulting from entropy maximization 
cannot depend explicitly on the spectral details of $\kappa _{\nu}$ and $\sigma _{\nu}$, which must be wrong in general.
A critical discussion of the maximum entropy production closure was already given by
\cite{Struchtrup1998}; he has shown that only a large number of moments generalized to higher powers in frequency
up to order $\nu ^4$, are able to reproduce the correct result in the weak nonequilibrium case. Consequently, despite of
its ostensible mathematical advantages, we propose to reject the maximum entropy closure for the moment expansion of
radiative heat transfer. A physically superior method based on the entropy production rate will be discussed in the next subsection.

\subsection{Minimum Entropy Production Rate Closure}
\label{MinEPSection}
As mentioned, \cite{Kohler1948} has proven that a minimum entropy production rate principle holds
for the linearized BTE. The application of this principle to moment expansions has been shown by \cite{Christen2009} for
the photon gas and by \cite{Christen2010} for a gas of independent electrons. The formal procedure is fully analogous to the
maximum entropy closure, but the functional to be minimized is in this case the {\em total entropy production rate}, which consist of
two parts associated with the radiation field, i.e., the photon gas, and with the LTE matter. The latter acts as a thermal
equilibrium bath. The two success factors of the application of this closure to radiative transfer
are first that the RTE is linear not only near equilibrium but in the whole range of $I_{\nu}$ (or $f_{\nu}$) values, and secondly
that the entropy expression Eq.
(\ref{radiationentropy}) is valid also far from equilibrium (cf. \cite{LandauLifshitz}).\\
In order ¦to derive the expression for the entropy production rate, $\dot S$, one
can consider separately the two partial (and spatially local) rates $\dot S_{\rm rad}$ and $\dot S _{\rm m}$ of the radiation
and the medium, respectively (cf.
\cite{Struchtrup1998}). $\dot S_{\rm rad}$ is obtained from the time-derivative of Eq. (\ref{radiationentropy}), use of Eq. (\ref{RTE}),
and writing the result in the form
$ \partial _{t} S_{\rm rad} + {\bf  \nabla \cdot J}_{S}= \dot S_{\rm rad} $ with
\begin{equation}
 \dot S_{\rm rad} [I_{\nu}] = -k_{B}\int d\nu \, d\Omega \frac{1}{h\nu} \ln \left(\frac{n_{\nu}}{1+n_{\nu}}
\right) \mathcal{ L}(B_{\nu}-I_{\nu}) \;\; ,
\label{radiationentropyproduction2}
\end{equation}
where $n_{\nu}$ is given by Eq. (\ref{photonnumber}).
${\bf  J}_{S}$ is the entropy current density, which is not of further interest in the following.
The entropy production rate of the LTE matter, $\dot S _{\rm mat} $, can be derived from the fact that
the matter can be considered locally as an equilibrium bath with temperature $T({\bf x})$.
Energy conservation implies that $W$ in Eq. (\ref{heatbalance}) is related to the
radiation power density in Eq. (\ref{Etdef}) by $W=-P_{E}$.
The entropy production rate (associated with radiation) in the local heat bath is thus $\dot S _{\rm mat} = W/T= -P_{E}/T$. Equation
(\ref{BoseEinstein}) implies $h\nu /k_{B}T = \ln (1+1/n_{\nu}^{(eq)})$, and one obtains  
\begin{equation}
 \dot S_{\rm mat} [I_{\nu}]= -k_{B}\int d\nu \, d\Omega \frac{1}{h\nu} \ln \left(\frac{1+n_{\nu}^{(eq)}}{n_{\nu}^{(eq)}}
\right) \mathcal{ L}(B_{\nu}-I_{\nu}) \;\; . 
\label{matterentropyproduction}
\end{equation}
The total entropy production rate $\dot S =\dot S_{\rm rad} + \dot S_{\rm mat} $ is
\begin{equation}
 \dot S [I_{\nu}]= \int _{0}^{\infty}d\nu \, \dot S_{\nu}
 =-k_{B}\int d\nu \, d\Omega \frac{1}{h\nu} \ln \left(\frac{n_{\nu}(1+n_{\nu}^{(eq)})}{n_{\nu}^{(eq)}(1+n_{\nu})}
\right) \mathcal{ L}(B_{\nu}-I_{\nu}) \;\; . 
\label{totalentropyproduction}
\end{equation}
The closure receipt prescribes to minimize $\dot S[I_{\nu}]$ by varying $I_{\nu}$ subject to the constraints that the moments
$E$, ${\bf F}$, ... etc.
are fixed. The solution $I_{\nu}$ of this constrained optimization problem depends on the values $E$, ${\bf F}$, ... . The number $N$
of moments to be taken into account is in principle arbitrary, but
we still restrict the discussion to $E$ and ${\bf F}$. After introducing
Lagrange parameters $\lambda _{E}$ and ${\bm \lambda _{F}}$, one has to solve
\begin{equation}
\delta _{I_{\nu}} \Biggl[ \dot S [I_{\nu}] - \lambda _{E} \left(  E -  \frac{1}{c}\int d\nu d\Omega \, I_{\nu} \right) -
 {\bm \lambda _{F}}\cdot \left(  {\bf F} -  \frac{1}{c} \int d\nu d\Omega \, {\bf \Omega} \,  I_{\nu} \right) \Biggr]=0\;\; \;\; , 
\label{MinEP}
\end{equation}
where $\delta _{I_{\nu}}$ denotes the variation with respect to $I_{\nu }$. The solution of
this minimization problem provides the nonequilibrium state $I_{\nu}$.\\
%

\section{Effective Transport Coefficients}
\label{Results}
We will now calculate the effective transport coefficients $\kappa _{E}^{\rm (eff)}$, $\kappa _{F}^{\rm (eff)}$,
and the Eddington factor $\chi $ with the help of the entropy production rate minimization closure.
We assume ${\bf F}= (0,0,F)$ in $x_{3}$-direction, use spherical coordinates $(\theta, \phi)$ in $\Omega$-space, such that
$I_{\nu}$ is independent of the azimuth angle $\phi$. For simplicity, we consider isotropic scattering with
$p({\bf \Omega}, {\bf \tilde \Omega})=1$,
although it is straightforward to consider general randomly oriented
scatterers with the phase function $p_{\nu}$ being a series in terms of Legendre polynomials $P_{n} (\mu)$.
Here, we introduced the abbreviation $\mu = \cos (\theta)$. With $d{\bf \Omega} = 2\pi \sin (\theta) d\theta = -2\pi d\mu$,
the linear operator $\mathcal{L} $, acting on a function $\varphi _{\nu} (\mu )$, can be written as
\begin{equation}
\mathcal{L} \varphi_{\nu }= 
\kappa _{\nu} \varphi _{\nu }(\mu) +\sigma _{\nu} \left( \varphi_{\nu } (\mu) -  \frac{1}{2}\int _{-1}^{1}
d\tilde \mu\,\varphi_{\nu } (\tilde \mu) \right) \;\;,
\label{linearmu}
\end{equation}
which has an eigenvalue $\kappa _{\nu}$ with eigenfunction $P_{0}(\mu)$ and (degenerated)
eigenvalues $\kappa _{\nu} +\sigma _{\nu}$ for all higher order Legendre polynomials $P_{n}(\mu), \;\; n= 1,2, ...\; $. 
In the following two subsections we focus first on limit cases that can be analytically solved, namely radiation
near equilibrium (leading order in $E - E^{(eq)}$ and $F$), and the emission limit
(leading order in $E $, while $0\leq F\leq E$). In the remaining subsections
the general behavior obtained from numerical solutions and a few mathematically relevant issues will be discussed.

\subsection{Radiation Near Equilibrium}
\label{NearEquilibrium}
Radiation at thermodynamic equilibrium obeys $I_{\nu}=B_{\nu}$ and $F=0$. Near equilibrium, or weak nonequilibrium,
refers to linear order in the deviation $\delta I_{\nu}= I_{\nu}-B_{\nu} $. Higher order corrections of the moments
$E=E^{(eq)}+\delta E$ and ${F}=\delta {F}$ are neglected. Because the stress tensor is
an even function of $\delta I_{\nu}$, $\chi = 1/3$ remains still valid in the linear nonequilibrium region (except for
the singular case of Auer's VEF with $j=1$).
We will now show that, in contrast to the entropy maximization closure, the entropy production minimization closure yields the
correct Rosseland radiation transport coefficients (cf. \cite{Christen2009}).\\
For isotropic scattering it is sufficient to take into account the first two Legendre polynomials, $1$ and $\mu $:
$\delta I_{\nu} =c^{(0)}_{\nu}+c^{(1)}_{\nu}\mu$, with $\mu $-independent $c^{(0,1)}_{\nu}$ that must be determined. Equations
(\ref{Edef}) and (\ref{Fdef}) yield
\begin{eqnarray}
\delta E_{\nu} & = & \frac{2\pi}{c}\int _{-1}^{1}d\mu \, (c^{(0)}_{\nu}+c^{(1)}_{\nu}\mu) =\frac{4\pi}{c}c^{(0)}_{\nu} \label{deltaE} \\
\delta F_{\nu} & = & \frac{2\pi}{c}\int _{-1}^{1}d\mu \, (c^{(0)}_{\nu}+c^{(1)}_{\nu}\mu)\mu =\frac{4\pi}{3c}c^{(1)}_{\nu} \label{deltaF} 
\end{eqnarray}
and from Eq. (\ref{totalentropyproduction})
\begin{equation}
\dot S_{\nu} = \frac{2k_{B}\pi c^2}{h^2\nu^4n_{\nu}^{(eq)}(1+n_{\nu}^{(eq)})}\left(\kappa _{\nu}(c^{(0)}_{\nu})^{2}+
\frac{1}{3}(\kappa _{\nu}+\sigma _{\nu}) (c^{(1)}_{\nu})^{2}  \right) \;\;.
\label{dotSnu}
\end{equation}
Minimization of $\dot S_{\nu}$ with respect to $c^{(0,1)}_{\nu}$ under constraints $\delta E = \int d\nu \delta E_{\nu}$ and
$\delta F = \int d\nu \delta F_{\nu}$ leads to
\begin{eqnarray}
c^{(0)}_{\nu} & = &  \frac{c\nu ^{4} \partial_{\nu}n_{\nu}^{(eq)}} {4\pi \kappa _{\nu} \int d\nu \,
\nu ^{4} \kappa _{\nu}^{-1} \partial_{\nu}n_{\nu}^{(eq)}}\delta E \label{c0} \\
c^{(1)}_{\nu} & = &  \frac{3c\nu ^{4} \partial_{\nu}n_{\nu}^{(eq)}} {4\pi (\kappa _{\nu} +\sigma_{\nu}) \int d\nu \,
\nu ^{4} (\kappa _{\nu}+\sigma_{\nu})^{-1} \partial_{\nu}n_{\nu}^{(eq)}}\delta F  \;\; ,  \label{c1} 
\end{eqnarray}
where we made use of the relation $\partial_{\nu}n_{\nu}^{(eq)} = n_{\nu}^{(eq)}(1+n_{\nu}^{(eq)}) h/k_{B}T $. As
$\delta I_{\nu}$ is known to leading order in $\delta E$ and $\delta F$, the transport coefficients can be calculated.
One finds
\begin{eqnarray}
\kappa _{E}^{\rm (eff)} = \frac{2\pi}{c}\int d\nu d\mu\,  \,  \frac{\mathcal{ L}(\delta I_{\nu})}{\delta E} & = & 
\frac{4\pi}{c}\int d\nu \kappa _{\nu}\frac{c^{(0)}_{\nu}}{\delta E} = \langle \kappa _{\nu} \rangle _{\rm Ro}
\label{effopacE} \\
\kappa _{F}^{\rm (eff)}= \frac{2\pi}{c}\int d\nu d\mu\, \mu \, \frac{\mathcal{ L}(\delta I_{\nu})}{\delta F} & = & 
\frac{4\pi}{c}\int d\nu (\kappa _{\nu}+\sigma _{\nu})\frac{c^{(1)}_{\nu}}{3\delta F} =\langle \kappa _{\nu}+\sigma _{\nu}
\rangle _{\rm Ro} \;\; ,
\label{effopacF} 
\end{eqnarray}
hence the effective absorption coefficients are given by the Rosseland averages Eqs. (\ref{RosselandmeanE}) and (\ref{RosselandmeanF}).
Similarly, it is shown that $\Pi _{kl} = (E/3)\delta _{kl}$. This
proves that the minimum entropy production rate closure provides the
correct radiative transport coefficients near equilibrium. 
\subsection{Emission limit}
\label{Emission}
While the result of the previous subsection was expected due to the general
proof by \cite{Kohler1948}, the emission limit
is another analytically treatable case, which is, however, far from equilibrium. It is characterized by a photon density much smaller than
the equilibrium density, hence $I_{\nu}\ll B_{\nu}$, i.e., $E\ll E^{(eq)}$, i.e., emission strongly predominates absorption.
To leading order in $n_{\nu}$, the entropy production rate becomes
\begin{equation}
\dot S_{\nu} = -2\pi k_{B} \int _{-1}^{1}d\mu \frac{\kappa _{\nu} B_{\nu}}{h\nu} \ln(n_{\nu})
\label{emission0}
\end{equation}
such that constrained optimization gives
\begin{equation}
I_{\nu} = \frac{2 k_{B}}{c}\frac{\nu^{2}\kappa _{\nu}}{\lambda_{E}+\lambda _{F} \mu } n_{\nu}^{(eq)} \;\; ,
\label{emission1}
\end{equation}
with Lagrange parameters $\lambda_{E}$ and $\lambda_{F}$.
The $\mu$-integration in Eqs. (\ref{Edef}) and (\ref{Fdef}) can be performed analytically, yielding 
\begin{eqnarray}
E & = & \frac{k_{B}\mathcal{T}(\kappa _{\nu})}{c^2\lambda _{F}} \ln \biggl( \frac{\lambda _{E} +\lambda _{F}}{\lambda _{E} -\lambda _{F}}\biggr)
\label{emission2} \\
F & = & \frac{k_{B}\mathcal{T}(\kappa _{\nu})}{c^2\lambda _{F}} \left( 2-
\frac{\lambda _{E}}{\lambda _{F}}\ln \biggl( \frac{\lambda _{E} +\lambda _{F}}{\lambda _{E} -\lambda _{F}}\biggr) \right)\;\; ,
\label{emission3} 
\end{eqnarray}
where we introduced
\begin{equation}
\mathcal{T}(\kappa _{\nu}) = 4 \pi \int _{0}^{\infty} d\nu\, \nu ^2\kappa _{\nu} n_{\nu}^{(eq)} \;\;.
\label{emission4}
\end{equation}
Up to leading order in $I_{\nu}$, one finds by performing the integration analogous to Eqs. (\ref{effopacE}) and (\ref{effopacF})
\begin{equation}
\kappa _{E} ^{(\rm eff)} =\langle \kappa _{\nu }\rangle _{\rm Pl} \;\;\;\;\;\; {\rm and} \;\;\;\;\;\; 
\kappa _{F} ^{(\rm eff)} = \frac{\mathcal{T}( \kappa _{\nu }(\kappa _{\nu} + \sigma _{\nu})) }{\mathcal{T}( \kappa _{\nu})} \;\;.
\label{emission5}
\end{equation}
As one expects, in the emission limit the effective absorption coefficients are Planck-like, i.e., a direct average rather than an average of the
inverse rates like Rosseland averages. 
The Eddington factor can be obtained from $\Pi _{33} = \chi E $ by calculating
\begin{equation}
\Pi _{33} = \frac{2\pi}{c}\int _{0}^{\infty} d\nu \int _{-1}^{1} d\mu\,\mu^2 I_{\nu} \;\; ,
\label{emission6a}
\end{equation}
which leads to
\begin{equation}
\chi (v)= - \frac{\lambda _{E}}{\lambda_{F}} v \;\;,
\label{emission6b}
\end{equation}
where the ratio of the Lagrange parameters, and thus also the VEF, depends only on $v=F/E$. This can be seen if one
divides Eq. (\ref{emission2}) by (\ref{emission3}).
For small $v$, the expansion of Eqs. (\ref{emission2}) and (\ref{emission3}) gives $\lambda _{E} /\lambda _{F} = - 1/3v$,
in accordance with the isotropic limit. In the free streaming limit $v\to 1$ from below, it holds
$\lambda _ {F} \to - \lambda _{E}$, which follows from $\ln (Z) = 2-\lambda_{E}\ln(Z)/\lambda _{F}$ with
$Z = (\lambda _E + \lambda _F)/ (\lambda _E - \lambda _F)$ obtained from equalizing (\ref{emission2}) with (\ref{emission3}).\\
For arbitrary $v$ the Eddington factor in the emission limit can easily be numerically
calculated by division of Eq. (\ref{emission2}) by Eq. (\ref{emission3}), and parameterizing $v$ and
$\chi $ with $\lambda_F/\lambda_E$. The result will be shown below in Fig. \ref{Chi_emission} a). It turns out that the difference
to other VEFs often used in literature is quantitatively weak.\\ 
While \cite{Christen2009} disregarded scattering, it is here included. For strong scattering
$\sigma_{\nu} \gg \kappa_{\nu}$, Eq. (\ref{emission5}) implies that the effective absorption coefficient $\kappa _{F} ^{(\rm eff)}$
of the radiation flux is
given by a special average of $\sigma _{\nu}$ where $\kappa _{\nu}$ enters in the weight function.
In particular, for frequencies where
$\kappa _{\nu}$ vanishes, there is no elastic scattering contribution to the average in this limit. This can be understood
by the absence of photons with this frequency in the emission limit.


\subsection{General Nonequilibrium Case}
\label{GeneralNonequilibrium}
\begin{figure}[H]	
\centering
{
\includegraphics[width=6.2cm]{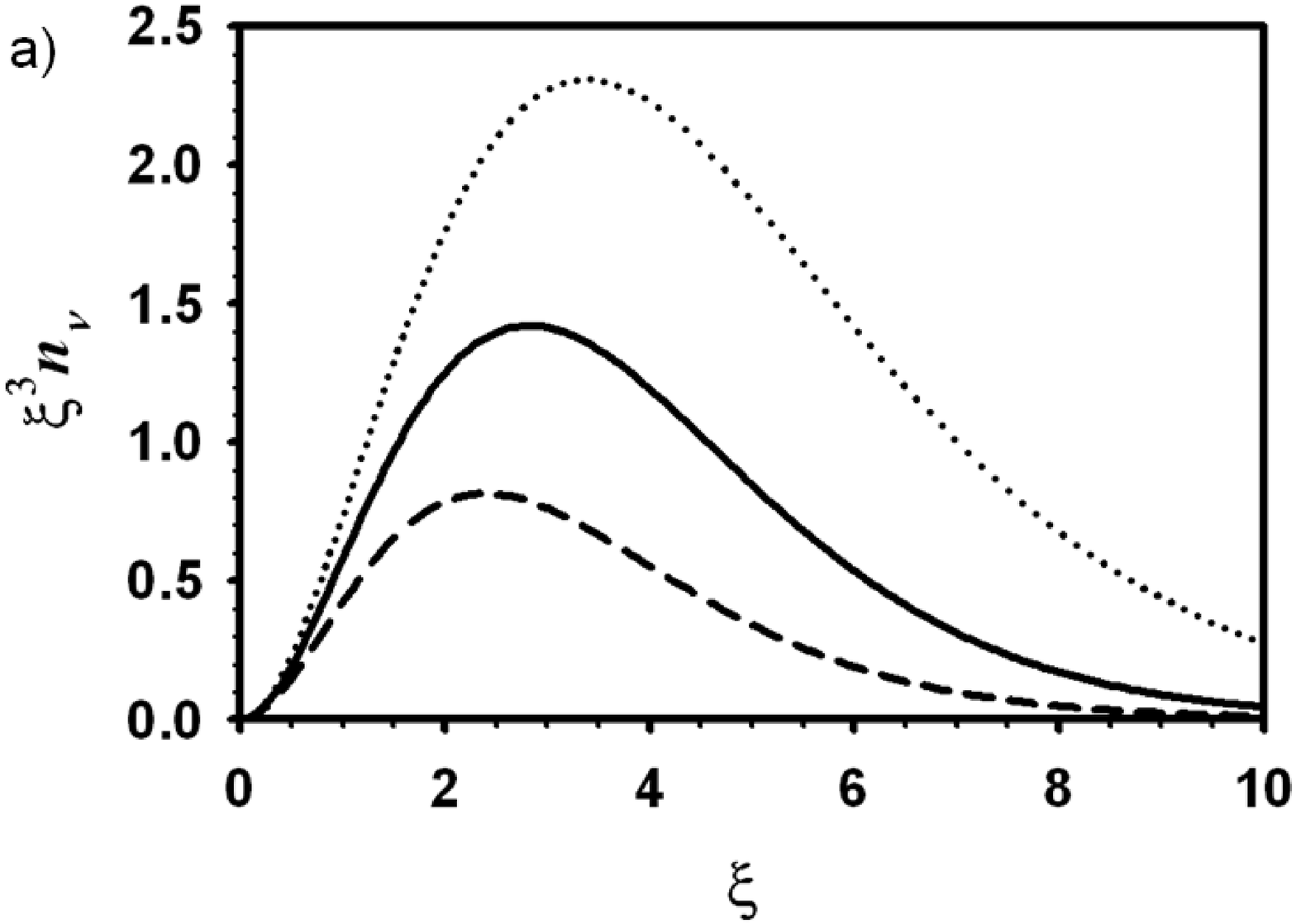}
\includegraphics[width=6.2cm]{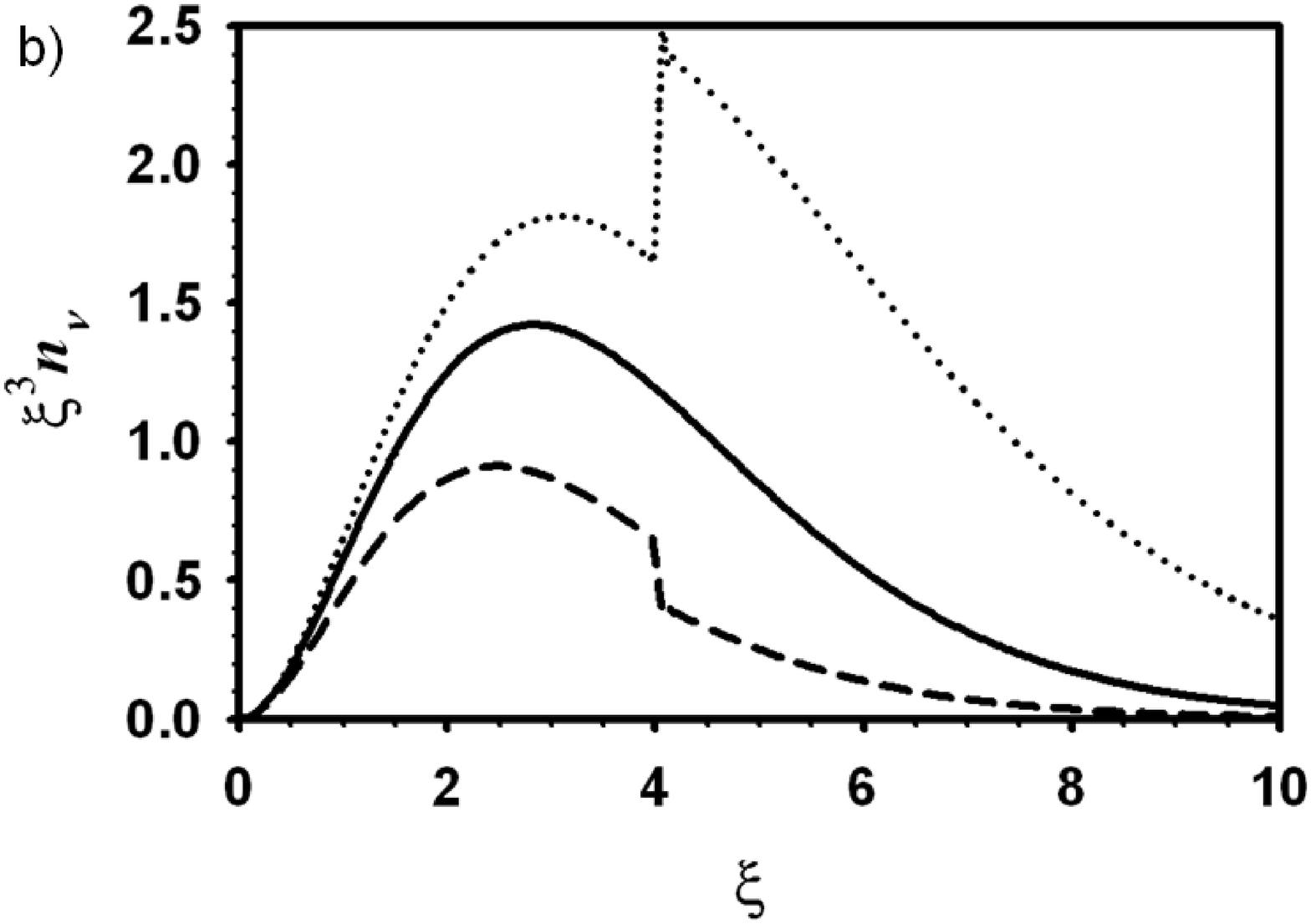}
}
\caption{
Nonequilibrium distribution ($\xi^3n_{\nu} \propto I_{\nu}$) as a function of $\xi =h\nu /k_{B}T$,
without scattering, for $F=0$ and $E=E^{(eq)}$ (solid),
$E=E^{(eq)}/2$ (dashed), and $E=2E^{(eq)}$ (dotted). a) gray matter; b) piecewise constant $\kappa $ with $\kappa _{\xi < 4} = 2 \kappa _{\xi > 4} $.}
\label{nvsxsi}
\end{figure}

The purpose of this subsection is to illustrate how the entropy production rate closure
treats strong nonequilibrium away from the just discussed limit cases.
For convenience, we introduce the dimensionless frequency $\xi =h\nu /k_{B}T$.
First, we consider gray-matter (frequency independent
$\kappa _{\nu}\equiv \kappa $) without scattering ($\sigma _{\nu} =0$). In Fig. \ref{nvsxsi} a) the quantity $\xi^3 n$, being
proportional to $I_{\nu} $, is plotted as a function of $\xi$ for $F=0$ and three values of $E$, namely $E=E^{(eq)}$,
$E=E^{(eq)}/2$, and $E=2E^{(eq)}$. The first case corresponds the thermal equilibrium with $I_{\nu}=B_{\nu}$, while
the others must have nonequilibrium populations of photon states. The results show that the energy unbalance is mainly due to
under- and overpopulation, respectively, and only to a small extent due to a shift of the frequency maximum.\\
Now, consider a non-gray medium, still without scattering, but with a frequency dependent
$\kappa _{\nu }$ as follows: $\kappa = 2 \kappa _{1}$ for $\xi < 4$, with constant $\kappa _{1}$, and $\kappa =  \kappa _{1}$ for $\xi > 4$.
The important property is that $\kappa _{\nu }$ is larger at low frequencies and smaller at high frequencies.
The resulting distribution function, in terms of $\xi ^{3} n$,
is shown in Fig. \ref{nvsxsi} b). For $E=E^{(eq)}$, the resulting distribution is of course still the Planck equilibrium distribution.
However, for larger (smaller) energy density the radiation density differs from the gray-matter case. In particular,
the distribution is directly influenced by the $\kappa _{\nu}$-spectrum. This behavior is not possible if one applies the
maximum entropy closure in the same framework of a single-band moment approximation. A qualitative
explanation of such behavior is as follows. Equilibration of the photon gas
is only possible via the interaction with matter. In frequency bands where the
interaction strength, $\kappa _{\nu}$, is larger ($\xi < 4$), the nonequilibrium distribution is pulled closer to the equilibrium
distribution than for frequencies with smaller $\kappa _{\nu}$.
This simple argument explains qualitatively the principal behavior associated with entropy production rate
principles: the strength of the irreversible processes determines the distance from thermal equilibrium in the presence of a
stationary constraint pushing a system out of equilibrium.

%
\begin{figure}[H]	
\centering
{
\includegraphics[width=6.2cm]{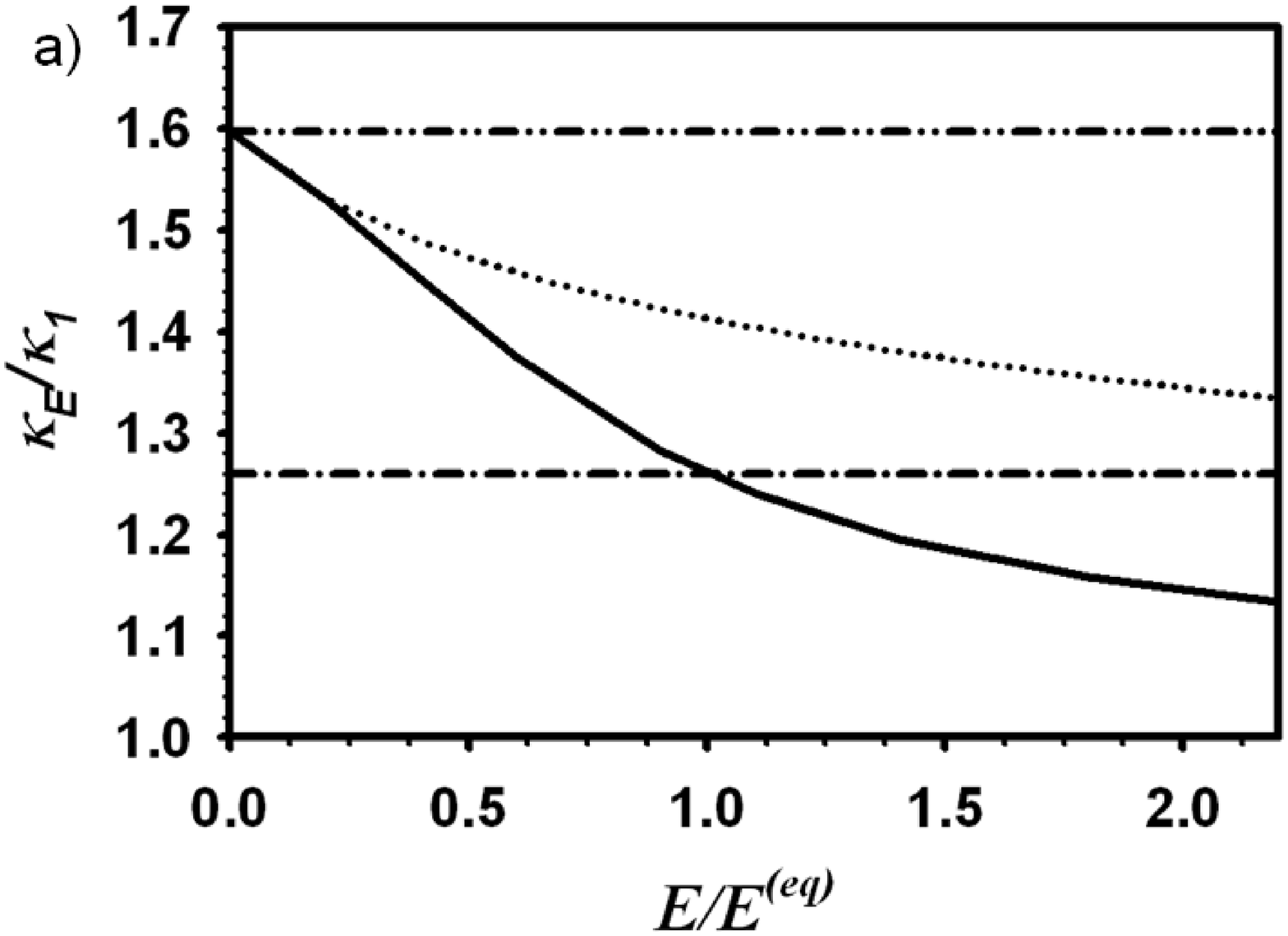}
\includegraphics[width=6.2cm]{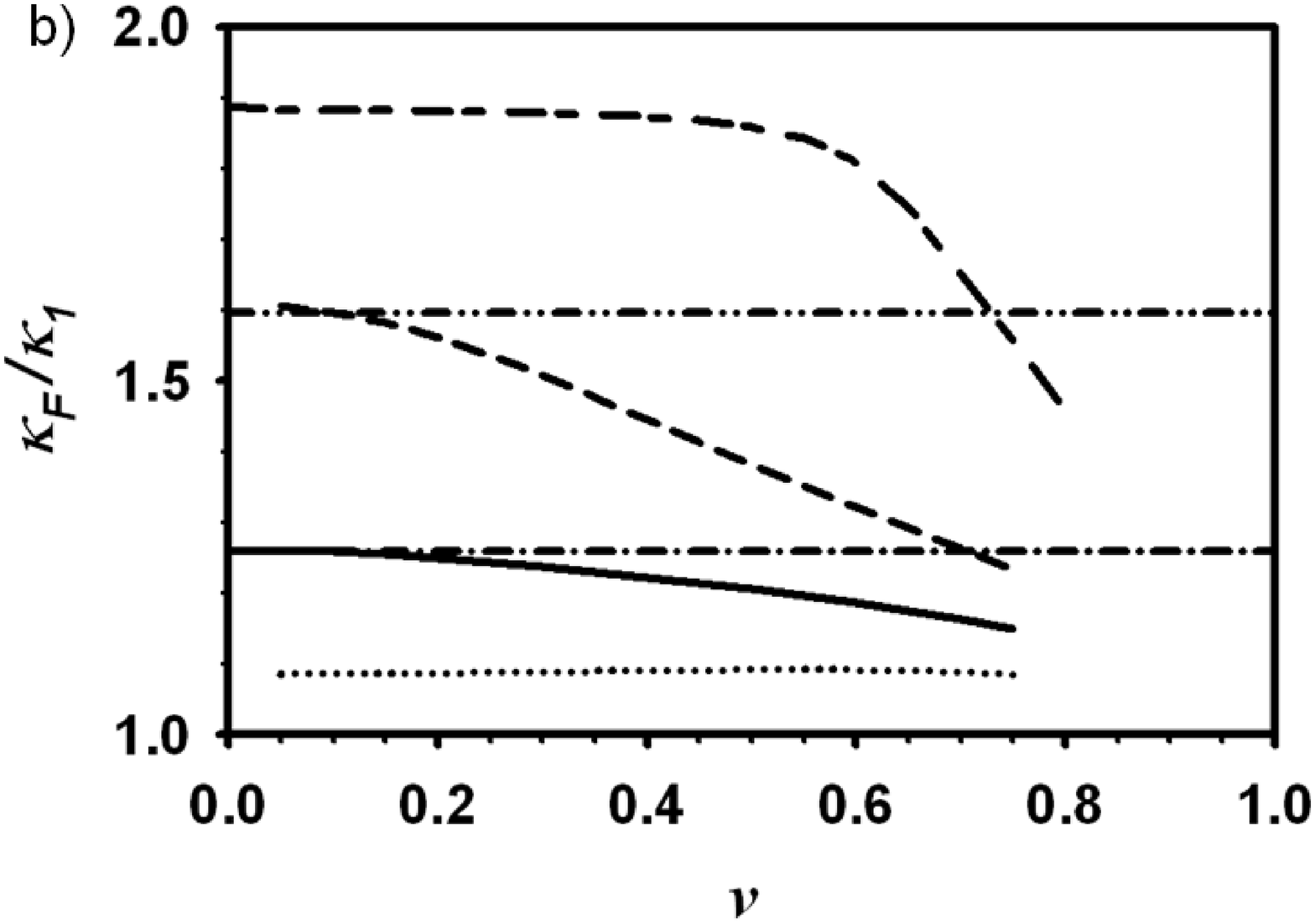}
}
\caption{
a) Effective absorption coefficients for $E$ as a function of $E$ for $F=0$, with the same spectrum as for
Fig. \ref{nvsxsi} b). Dashed-dotted: Rosseland mean; dashed-double-dotted: Planck mean;
solid: entropy production rate closure (correct at $E=E^{(eq)}$); dotted: entropy closure (wrong at $E=E^{(eq)}$).
b) Effective absorption coefficients for ${\bf F}$ as a function of $v=F/E$
for different $E$-values (dotted: $E/E^{(eq)}=2$;
solid $E/E^{(eq)}=1$; dashed: $E/E^{(eq)}=0.5$; short-long dashed: $E/E^{(eq)}=0.05$). Dashed-dotted and dashed-double dotted as in a).}
\label{kappaEandF}
\end{figure}

\begin{figure}[H]	
\centering
{
\includegraphics[width=6.2cm]{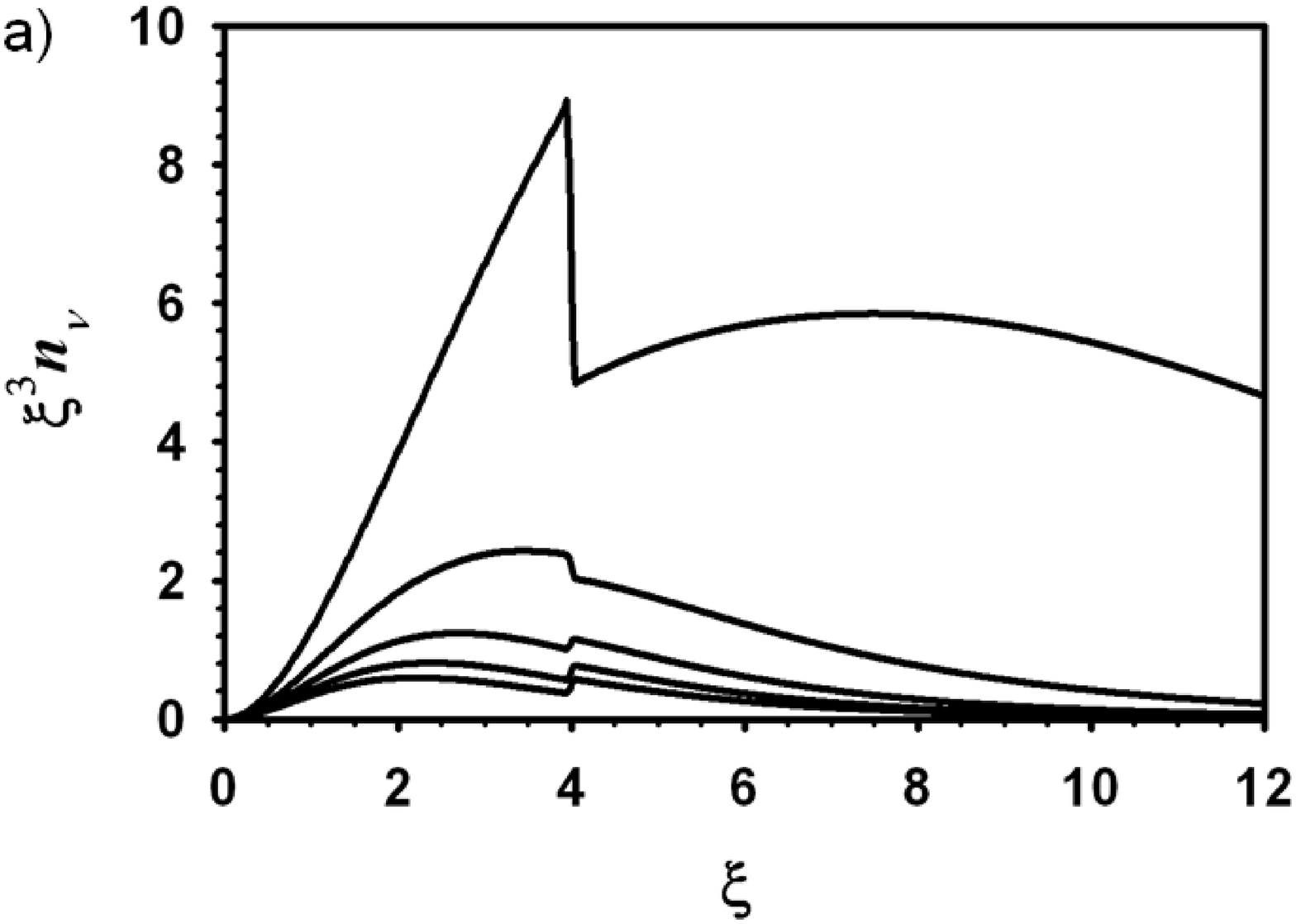}
\includegraphics[width=6.2cm]{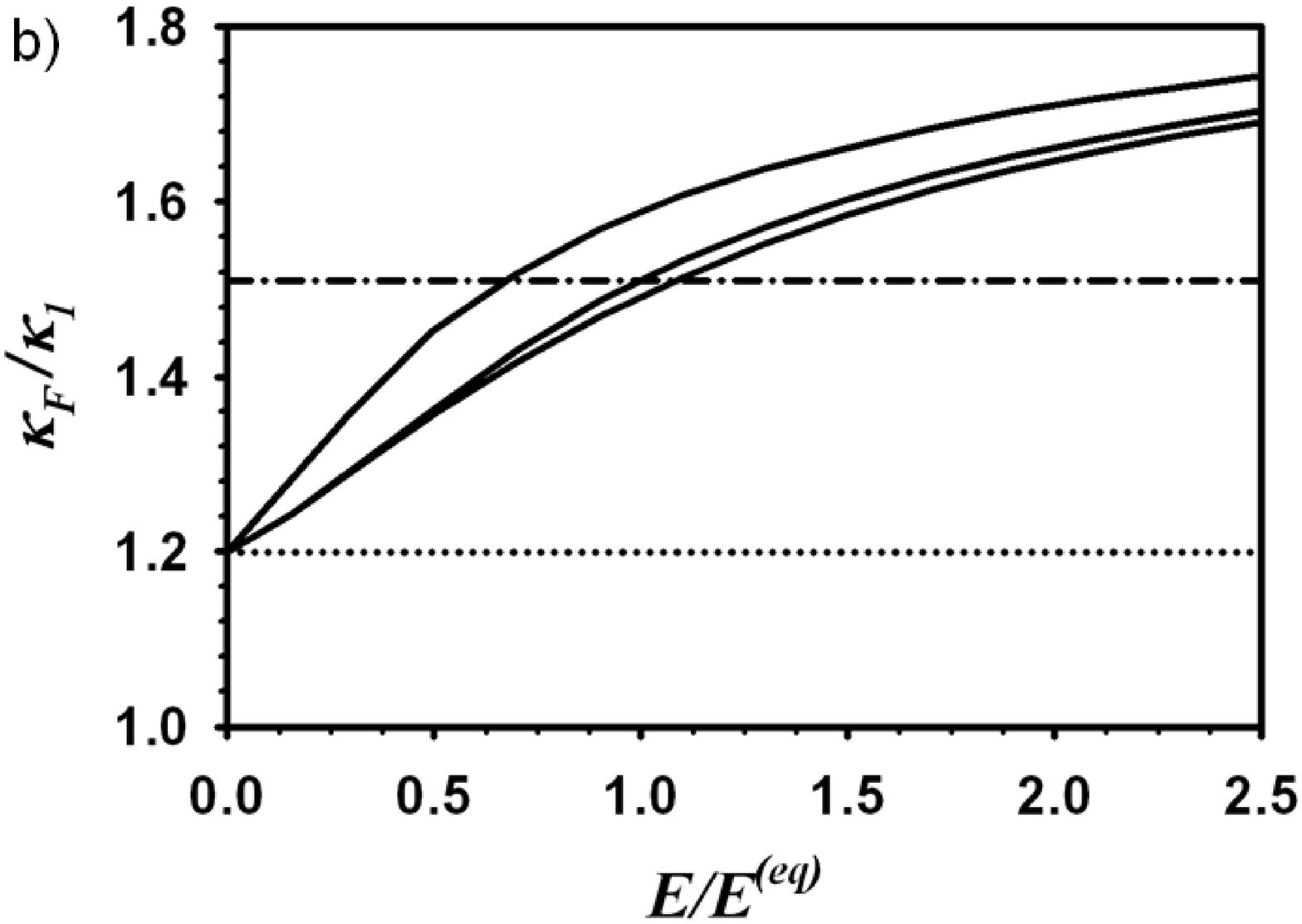}
}
\caption{
a) Nonequilibrium distribution ($\xi^3n_{\nu} \propto I_{\nu}$) as a function of $\xi =h\nu /k_{B}T$,
for a medium with constant absorption $\kappa _{\nu}\equiv \kappa _{1}$ and piecewise constant scattering $\sigma _{\mu}$, with
$\sigma _{\xi < 4} = 0$, and $ \sigma  _{\xi > 4}=\kappa_{1}$. The different curves refer to different radiation directions of
$\mu = -1,\,  -0.5, \, 0, \, 0.5,\,  1$ (solid curves in ascending order) from photons counter-propagating to the mean drift ${\bf F}$ to photons in
${\bf F}$-direction. b) Effective absorption coefficients $\kappa _{F}^{\rm (eff)}$ as a function of $E/E^{(eq)}$ for $v=0.125, \, 0.25,\, 0.5$ (solid curves
in ascending order); dashed-dotted: Rosseland mean, dashed: emission mean of $\kappa _{F} ^{(\rm eff)}$.}
\label{sigmastep}
\end{figure}

Results for the effective absorption coefficients $\kappa _{E}^{\rm (eff)}$ and $\kappa _{F}^{\rm (eff)}$ are shown in Fig. \ref{kappaEandF}.
In Fig. \ref{kappaEandF} a) it is shown that the effective absorption coefficient $\kappa _{E}^{\rm (eff)}$ is equal to the Planck mean
($1.26\,\kappa _{1}$, dashed-double-dotted)
in the emission limit $E/E^{(eq)}\to 0$, and equal to the
Rosseland mean ($1.6\, \kappa _{1}$, dashed-dotted) near equilibrium $E=E^{(eq)}$, and eventually goes slowly to the high frequency
value $\kappa _{1}$ for large $E$. The effective absorption coefficient obtained from the maximum entropy closure is also plotted (dotted curve),
and although correct for $E/E^{(eq)}\to 0$, it is wrong at equilibrium $E=E^{(eq)}$.
For the present example the maximum entropy closure is strongly overestimating the values of
$\kappa _{E}^{\rm (eff)}$.\\
Figure \ref{kappaEandF} b) shows $\kappa _{E}^{\rm (eff)}$ as a function $v$, for various values of $E$. As at constant $E$,
increasing $v$ corresponds to a shift of the distribution towards higher frequencies in direction of ${\bf F}$, a decrease
of $\kappa _{E}^{\rm (eff)}$ must be expected, which is clearly observed in the figure.
%
In order to investigate the effect of scattering $\sigma _{\nu} \neq 0$, we consider the example of
gray absorbing matter, i.e., constant $\kappa _{\nu} \equiv \kappa_{1}$, having a frequency dependent scattering rate
$\sigma _{\xi < 4} = 0$ and $\sigma _{\xi > 4} = \kappa _{1}$. Scattering is only active
for large frequencies. The distribution $\xi ^{3}n_{\nu}$ of radiation with $E=2E^{(eq)}$,
with finite flux $v = 0.25$ for different directions $\mu = \cos(\theta)= -1,\, -0.5, \, 0, \, 0.5, \, 1$ is plotted in Fig. \ref{sigmastep} a).
Since the total energy of the photon gas is twice the equilibrium energy, 
the curves are centered around about twice the equilibrium distribution. As one expects, the states in forward direction ($\mu =1$)
have the highest population, while the states propagating against the mean flux ($\mu = -1$) have lowest population.
This behavior occurs, of course, also in the absence of scattering. One observes that scattering acts to decrease the
anisotropy of the distribution, as for $\xi > 4$ the curves are pulled towards the state with $\mu \approx 0$.
Hence, also the effect of elastic scattering to the distribution function can be understood in the framework of the entropy production,
namely by the tendency to push the state towards equilibrium with a strength related to the interaction with the LTE matter. 

The effective absorption coefficient $\kappa _{F}^{\rm (eff)}$ is shown in Fig. \ref{sigmastep} b) for various values of $v$; it is obvious that it must increase for
increasing $v$ and for increasing $E$. The Rosseland and Planck averages
of $\kappa _{\nu}+\sigma_{\nu}$ are given by $1.42 \, \kappa_{1}$ and $1.40 \, \kappa_{1}$, while the emission limit for $\kappa _{F} ^{(\rm eff)}$
given in Eq. (\ref{emission5}) is $1.20\, \kappa_{1}$.\\
The VEF will be discussed separately in the following subsection, because its behavior has not only quantitative physical, but also
important qualitative mathematical consequences.


\subsection{The Variable Eddington Factor and Critical Points}
\label{Hyperbolicity}
A detailed discussion of general mathematical properties and conventional closures is given by \cite{Levermore1996}.
A necessary condition for a closure method is existence and uniqueness of the solution.
It is well-known that convexity of a minimization problem is a crucial property in this context.
One should note that convexity of the entropy production rate in nonequilibrium situations is often introduced
as a presumption for further considerations rather than it is a proven property (cf. \cite{Martyushev2006}).
For the case without scattering, $\sigma _{\nu} \equiv 0$,
\cite{Christen2009} have shown that the entropy production rate (\ref{MinEP}) is strictly convex.
A discussion of convexity for a finite scattering rate goes beyond the purpose of this chapter.\\
Besides uniqueness of the solution, the moment equations should be of hyperbolic type, in order to come up with a physically
reasonable radiation model. It is an advantage of the entropy maximization closure
that uniqueness and hyperbolicity are fulfilled and are related to the convexity properties of the entropy (cf. \cite{Levermore1996}).
In the following, we provide some basics needed for understanding the problematic of hyperbolicity, its relation to the VEF and
the occurrence of critical points. The latter is practically relevant because it affects the modelling of 
the boundary conditions, particularly in the context of numerical simulations.
More details are provided by \cite{Koerner1992}, \cite{Smit1997}, and \cite{Pons2000}.\\

A list of the properties that a reasonable VEF must have (cf. \cite{Pomraning1982}) is:  $\chi (v=0)=1/3$, $\chi (v=1)=1$,
monotonously increasing $\chi (v)$, and the Schwarz inequality $v^2 \leq \chi (v) $. The latter
follows from the fact that $\chi $ and $v$ can be understood as averages of $\mu ^2$ and $\mu $, respectively,
with (positive) probability density $I_{\nu}(\mu)/E$. Hyperbolicity adds a further requirement to the list.
Equations (\ref{Enueq}) and (\ref{Fnueq}) form a set of quasilinear first order differential equations. For simplicity, we consider
a one dimensional position space\footnote{Momentum space remains three dimensional.}
with coordinate $x$ with $0\leq x\leq L$,
and variables $E\geq 0$ and $F$. In this case we redefine $F$, such that it can have either sign, $-E\leq F\leq E$. We assume 
flux in positive direction, $F\geq 0$, write the moment equations in the form
\begin{equation}
\frac{1}{c}\partial _{t}
\left( \begin{array}{ccc}
E  \\
F
\end{array}\right)
+
\left( \begin{array}{ccc}
0 & 1 \\
\partial _{E}(\chi E) &  E\partial _{F}\chi 
\end{array}\right)\partial _{x}
\left( \begin{array}{ccc}
E  \\
F
\end{array}\right)
=
\left( \begin{array}{ccc}
P_{E}  \\
P_{F}
\end{array}\right) \;\; .
\label{LPDE}
\end{equation}
For spatially constant $E$ and $F$, small disturbances of $\delta E$ and $\delta F$ must propagate with 
well-defined speed, implying real characteristic velocities. Those are given by the eigenvalues of the matrix that 
appears in the second term on the left hand side of Eq. (\ref{LPDE}) and which we denote by ${\bf M}$:
\begin{equation}
w _{\pm} =  \frac{{\rm Tr}\,{\bf M}}{2} \pm \sqrt{\frac{({{\rm Tr}\,{\bf M}})^{2}}{4}-{{\rm det}\,({\bf M}})} \;\; ,
\label{charvel}
\end{equation}
where "Tr" and "det" denote trace and determinant. Note that the $w_{\pm}$ are normalized to $c$, i.e.
$-1\leq w_{-}\leq w_{+}\leq 1$ must hold. Hyperbolicity refers to real eigenvalues $w_{\pm }$ and to the existence of two independent eigenvectors.
The condition for hyperbolicity reads $(\partial_{F}(\chi E))^2+4\partial_{E}(\chi E) > 0$.\\
Provided hyperbolicity is guaranteed, the sign of the velocities is an issue relevant for the boundary conditions.
Indeed, the boundary condition, say at $x=L$, can only have an effect on the state in the domain if at least one of
the characteristic velocities is negative. It is clear that a disturbance near equilibrium ($v=0$) propagates
in $\pm x$ direction since $w_{+} = -w_{-}$ due to mirror symmetry. Hence $w_{-}<0<w_{+}$ for sufficiently small $v$.
In this case boundary conditions to both boundaries $x=0$ and $x=L$ have to be applied as in a usual boundary value problem.
However, for finite $v$, reflection symmetry is broken and $w_{+}\neq -w_{-}$. It turns out, that for sufficiently large
$v$, either $w_{+}$ or $w_{-}$ can change sign. For positive $F$, we denote the value of $v$ where $w_{-}$
becomes positive by $v_{c}$. This is called a {\em critical point}
because ${\rm det} ({\bf M})=w_{+}w_{-}$ vanishes there. Beyond the critical point, all disturbances will propagate in positive direction,
and a boundary condition at $x=L$ is not to be applied. This can introduce a problem in numerical simulations with fixed predefined
boundary conditions.
The rough physical meaning of the critical point is the cross-over from diffusion dominated to streaming dominated radiation.
In the latter region it might be reasonable to improve the radiation model by involving higher order moments or
partial moments, for example by decomposing the moments in backward and forward propagating components $E_{\pm}$ and $F_{\pm}$
(cf. sect. 3.1 in \cite{Frank2007}).\\

\begin{figure}[H]	
\centering
{
\includegraphics[width=6.2cm]{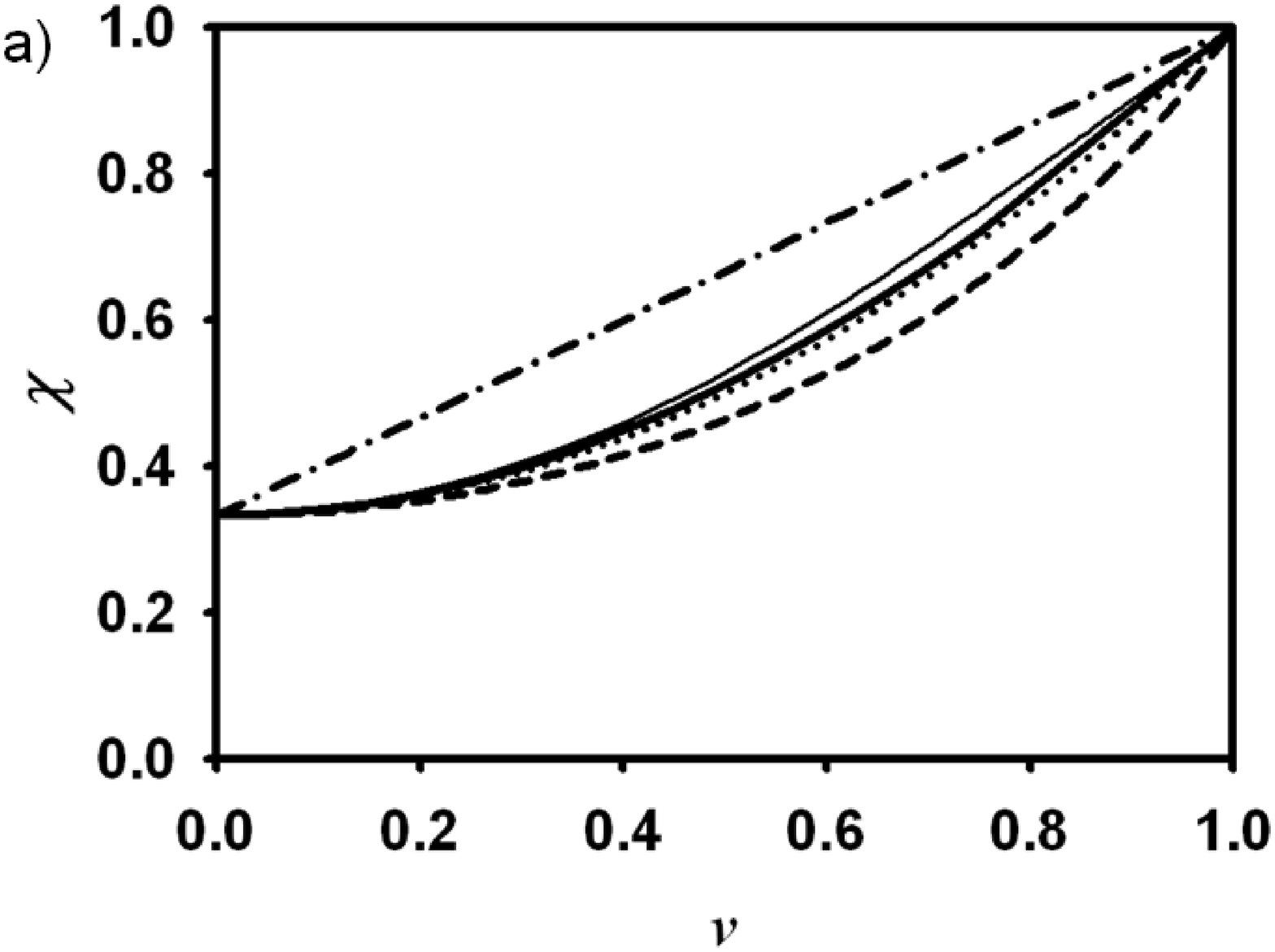}
\includegraphics[width=6.2cm]{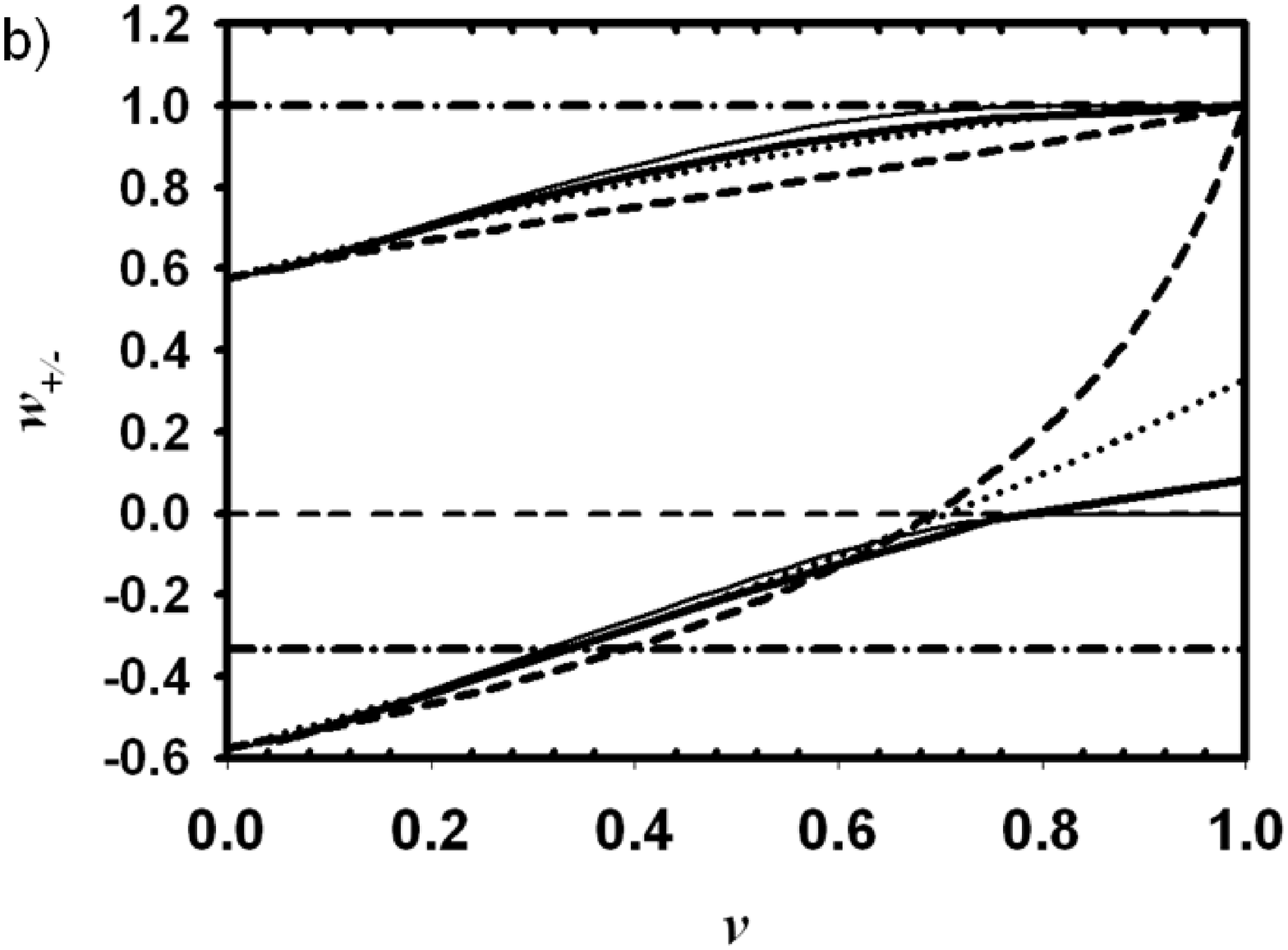}
}
\caption{
a) Eddington Factors $\chi $ versus $v$ and b) characteristic velocities $w_{\pm}$ (see Sect. \ref{Hyperbolicity})
for various cases. Minimum entropy production:
$E=E^{(eq)}$ (thick solid curve) and emission limit $E\ll E^{(eq)}$ (thin solid curve); maximum entropy (dashed);
Kershaw (dotted; $j=2$ in Eq. (\ref{VEFsimple})), and Auer (dashed-dotted; $j=1$ in Eq. (\ref{VEFsimple})).}
\label{Chi_emission}
\end{figure}

In Fig. \ref{Chi_emission} a), different VEFs are shown. All of them exhibit
the above mentioned properties, $\chi (v=0)= 1/3$, monotonous increase,
$\chi (v\to 1)=1$, and the Schwarz inequality $v^2\leq \chi $. 
In particular, the VEFs obtained from entropy production rate minimization
is shown for $E=E^{(eq)}$ for gray matter with
$\sigma_{\nu}\equiv 0$, as well as for the emission limit (cf. Eqs. (\ref{emission2}) and (\ref{emission3})).
Note that the latter $\chi (v)$ is a function of $v$ only and is
independent of the detailed properties of the absorption and scattering spectra.
The similarity of the differently defined VEFs, combined with the error done anyhow by the
two-moment approximation, makes it obvious that for practical purpose the simple Kershaw VEF ($j=2$) may serve as a sufficient approximation.\\
In Fig. \ref{Chi_emission} b) the characteristic velocities $w _{\pm}$ are plotted versus $v$ for the various VEFs discussed above. It turns out
that the VEF given by Eq. (\ref{VEFsimple}) has a critical point for $j>3/2$ given by $v_{c}=1/\sqrt[j]{2(j-1)}$, and
that there is a minimum $v_{c}$ value of $0.63$ at $j=3.16$. The VEF by Kershaw and maximum entropy have
$v_{c}=1/\sqrt{2}$ and $v_{c}=2\sqrt{3}/5$, respectively. Also the VEF associated with the entropy production rate has generally
a critical point, which depends on $E$. One has to expect a typical value of $v_{c}\approx 2/3$. For the VEF (\ref{VEFsimple})
with $j=1$ a critical point will not appear. In the framework of numerical simulations, this advantage can outweigh in certain 
situations the disadvantage of the erroneous anisotropy in the $v\to 0$ limit.\\

%
%

\section{Boundary Conditions}
\label{BoundaryConditions}
In order to solve the moment equations, initial and boundary conditions are required. While the definition
of initial conditions are usually unproblematic, the definition of boundary conditions is not straight-forward and deserves some
remarks. In the sequel we will consider boundaries where the characteristic velocities are such that boundary conditions are needed.
But note that the other case where boundary conditions are obsolete can also be important, for example in stellar physics where, beyond a certain distance
from a star, freely streaming radiation completely escapes into the vacuum.\\
The mathematically general boundary condition for the two-moment model is of the form 
\begin{equation}
aE+b{\bf \hat n}\cdot{\bf F} = \Gamma \;\; ,
\label{BCgeneral1}
\end{equation}
with the surface normal ${\bf \hat n}$, and
where the coefficients $a$, $b$, and the inhomogeneity $\Gamma$ must be determined in principle from
Eq.(\ref{boundaryconditionRTE}). There is a certain ambiguity to do this (cf. \cite{DuderstadtMartin1979})
and thus a number of different boundary conditions exist in the literature (cf. \cite{Su2000}).\\
There may be simple cases where one can either apply Dirichlet boundary conditions $E({\bf x}_{\rm w})=E_{\rm w}$ to $E$,
where $E_{\rm w}$ is the equilibrium value associated with the (local) wall temperature, and/or
homogeneous Neumann boundary conditions to ${\bf F}$, $({\bf \hat n}\cdot{\bf \nabla}) {\bf F }=0$, at ${\bf x}_{\rm w}$.
This approach may work,
if the boundaries do not significantly influence the physics in the region of interest, e.g., in the case where cold
absorbing boundaries are far from a hot radiating object under investigation. It can also be convenient to include in the simulation,
instead of using boundary conditions, the solid bulk material that forms the surface, and to describe it by its
$\kappa _{\nu}$ and $\sigma _{\nu}$. In the next section an example of this kind will be discussed.
If necessary, thermal equilibrium boundary conditions deep inside the solid may be assumed.
In this way, it is also possible to analytically calculate the Stefan-Boltzmann radiation law for
a plane sandwich structure (hot solid body)-(vacuum gap)-(cold solid body), if an Eddington factor (\ref{VEFsimple})
with $j=1$ is used and the solids are thick opaque gray bodies.\\
In general, however, one would like to have physically reasonable boundary conditions at a surface characterized by
Eq. (\ref{boundaryconditionRTE}). For engineering applications, often boundary conditions  by \cite{Marshak1947}
are used. In the following, we sketch the principle how these boundary conditions can be derived (cf. \cite{Bayazitoglu1979}).
For other types, like Mark or modified Milne boundary conditions (cf. \cite{Su2000}).
Let the coordinate $x\geq 0$ be normal to the surface at $x=0$,
and ask for the relation between the normal flux $F$, $E$, and $E_{\rm w}^{(eq)}$ at $x= 0$.
The ${\bf F}$-components tangential to the boundary are assumed to vanish, and diffusive reflection with $r ({\bf x_{w}},{\bf \Omega}
,{\bf \tilde \Omega} )= r/\pi$ with $r=1-\epsilon$ is considered. In terms of moments, the radiation field is given by
\begin{equation}
I_{\nu}=\frac{c}{4 \pi}\left(E_{\nu} P_{0}(\mu) +3F_{\nu } P_{1}(\mu) + \frac{5}{2}(3\Pi_{\nu,\, 11}-E_{\nu})P_{2}(\mu) + ...\right) \;\;,
\label{Iexpansion}
\end{equation}
with Legendre polynomials $P_{0}=1$, $P_{1}=\mu$, $P_{2}=(3\mu ^2-1)/2$, and where $\Pi_{11}=\chi E$.
The exact solution contains also higher order Legendre
polynomials, as indicated by the dots. The boundary condition
(\ref{boundaryconditionRTE}) can be written as
\begin{equation}
I_{\nu}(\mu \geq 0) = \epsilon B_{\nu} + 2 r \int _{-1}^{0}d\tilde \mu\, \mid \tilde \mu \mid I_{\nu} (\tilde \mu) \;\;.
\label{boundaryconditionRTEsimple}
\end{equation}
By using Eq. (\ref{Iexpansion}), the integral can be calculated, such that the right hand side of Eq. (\ref{boundaryconditionRTEsimple})
becomes a constant with respect to $\mu$, while the left hand side is, according to Eq. (\ref{Iexpansion}), a function of $\mu$
defined for $0\leq \mu \leq 1$. In order to obtain the required relation between $F$ and $E$,
one has to multiply Eq. (\ref{boundaryconditionRTEsimple}) with a weight function $h(\mu)$ and integrate over $\mu $ from 0 to 1.
The above mentioned ambiguity lies in the freedom of choice of $h(\mu)$. \cite{Marshak1947} selected $h=P_{1}$.
Provided $P_{n}$ for $n>3$ are neglected in Eq. (\ref{Iexpansion}), the integration leads to
\begin{equation}
F = \frac{\epsilon }{2(2-\epsilon)} \left( E_{\rm w} -\frac{(3E+15\,\Pi _{11})}{8} \right) \;\;.
\label{BCgeneral2}
\end{equation}
If higher order moments are to be considered, additional projections have to be
performed, in analogy to the procedure reported by \cite{Bayazitoglu1979} for the P-3 approximation.\footnote{Note that neither the series (\ref{Iexpansion}) stops after
the $N$'th moment (even not for the P-N approximation, cf. \cite{Cullen2001}),
nor all higher order coefficients drop out after projection of Eq. (\ref{boundaryconditionRTEsimple}) on $P_{n}$.
A general discussion, however, goes beyond this chapter in will be published elsewhere.}
For isotropic radiation with $\chi = 1/3$, or $\Pi_{11}=E/3$,  the prefactor of $E$ becomes unity and Eq. (\ref{BCgeneral2})
reduces to the well-known P-1-Marshak boundary condition. In the transparent limit 
with $\chi =1$, the prefactor becomes $9/4$.\\
For the simple case of two parallel plane plates ($\epsilon
=1$) with temperatures associated with $E_{{\rm w},1}$ and $E_{{\rm w}, 2}<E_{{\rm w}, 1}$, and separated by a vacuum gap,
both moments $E$ and $F$ are spatially constant and the Stefan-Boltzmann law $F=(E_{1}^{(eq)}-E_{2}^{(eq)})/4$ is
recovered. But note that the energy density $E$ between the plates is not equal to the expected
average of $E_{1}^{(eq)}$ and $E_{2}^{(eq)}$,
which is an artifact of the two-moment approximation with VEF.


\section{A Simulation Example: Electric Arc Radiation}
\label{Arc}
\begin{figure}[H]	
\centering
{
\includegraphics[width=6.2cm]{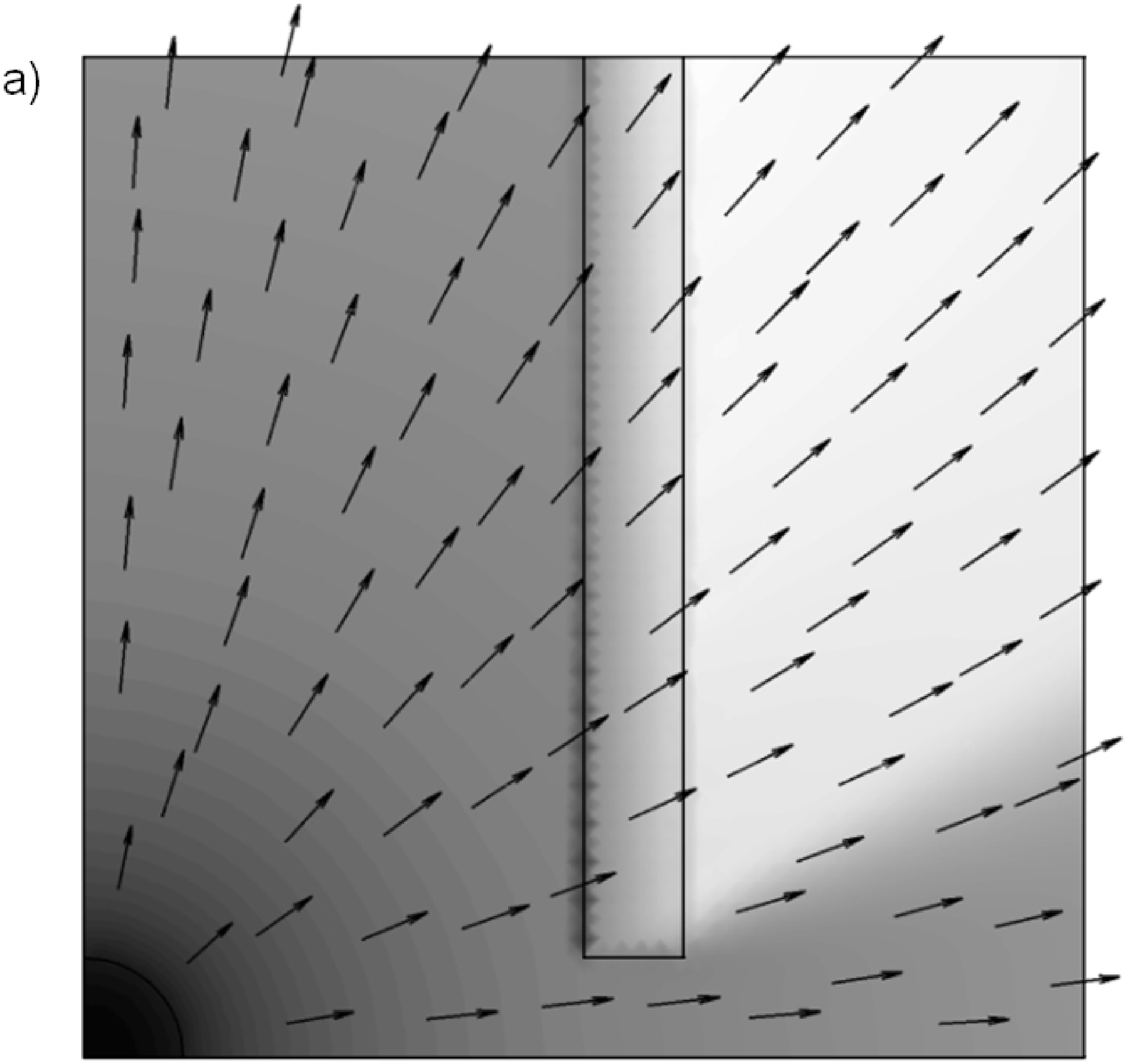}
\includegraphics[width=6.2cm]{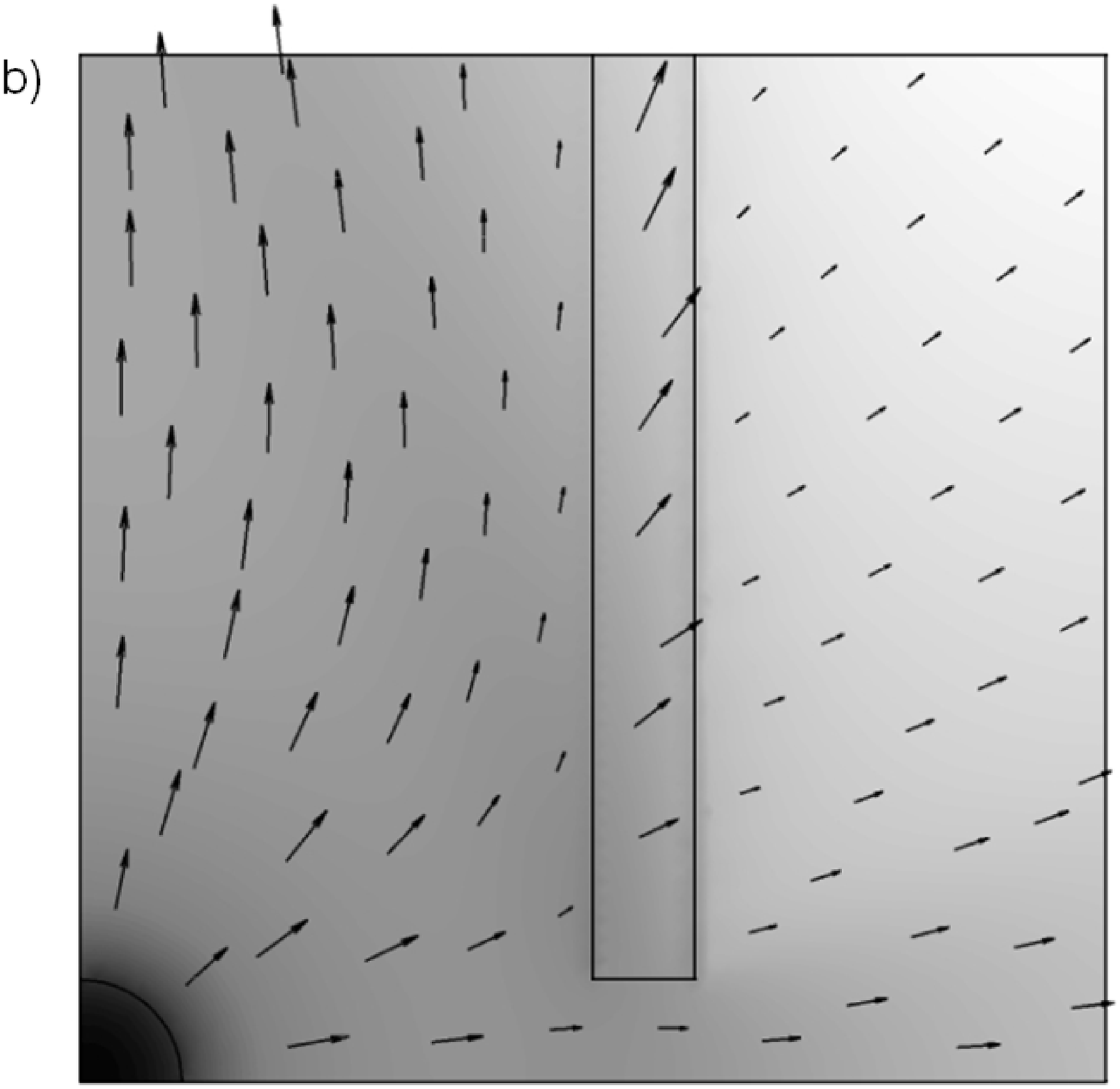}
}
\caption{
Illustrative simulations of the moment equations with FLUENT$^{\textregistered}$ for a cylindrical electrical arc (radius $1$ cm, temperature $10'000$ K,
$\kappa _{E} ^{\rm (eff)} = \kappa _{F} ^{\rm (eff)} = 1/m$) in a gas (ambient temperature $300$ K,
$\kappa _{E} ^{\rm (eff)} = \kappa _{F} ^{\rm (eff)} = 5/m$).
A solid wall (a): only absorbing with
$\kappa _{E} ^{\rm (eff)} = \kappa _{F} ^{\rm (eff)} \equiv 500/m $; b) wall with scattering coefficient
$\kappa _{E} ^{\rm (eff)} = 5/m$, $\kappa _{F} ^{\rm (eff)} \equiv 500/m $) with a slit in front of the arc
focuses the radiation towards a wall. Surface plot for $E$ (dark: large, bright: small, logarithmic scale);
arrows for ${\bf v}$ (not ${\bf F}$!). Only one quadrant of the symmetric arrangement is show.}
\label{arc2d}
\end{figure}

\begin{figure}[H]	
\centering
{
\includegraphics[width=6.2cm]{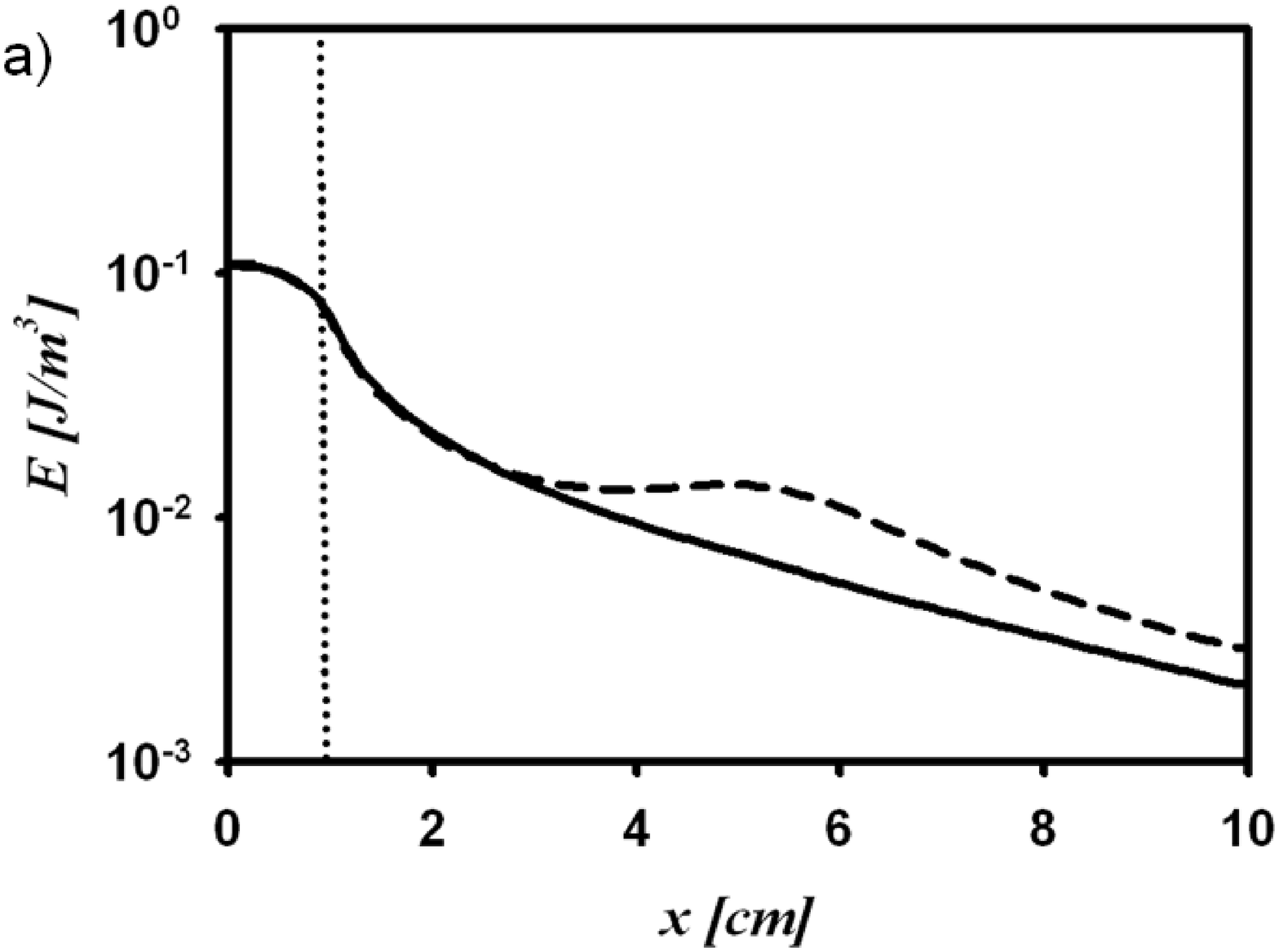}
\includegraphics[width=6.2cm]{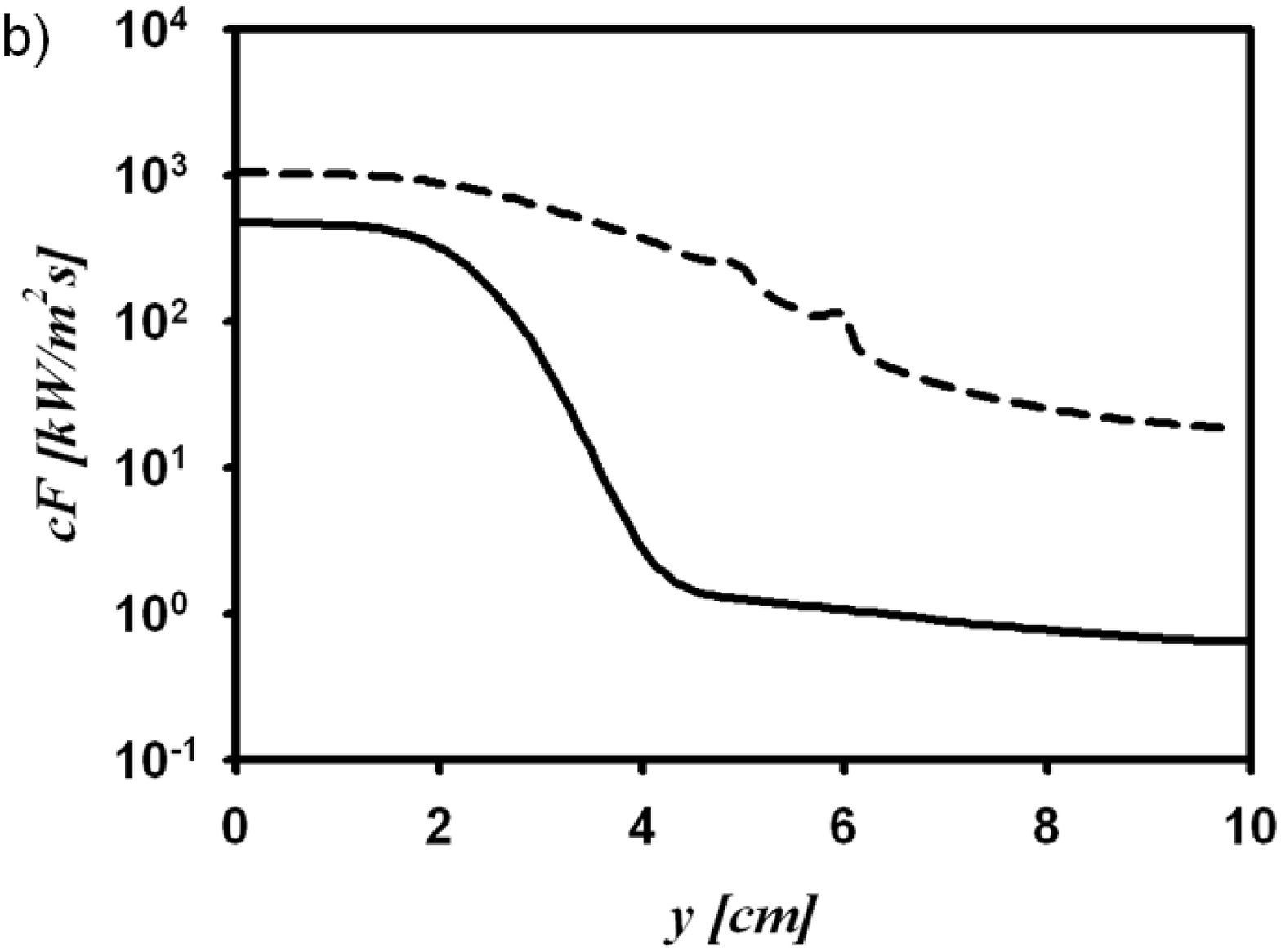}
}
\caption{a) Energy density along the $x$-axis (arc center at $x=0$) and b) power flux along the
screen ($x=10\, cm$) for the two cases Fig. \ref{arc2d} a) (solid)
and Fig. \ref{arc2d} b) (dashed).}
\label{arcEandF}
\end{figure}

The two-moment approximation will now be illustrated for the example of an electric arc.
The extreme complexity of the full radiation hydrodynamics is obvious.
Besides transonic and turbulent gas dynamics, which is likely supplemented with side effects like mass
ablation and electrode erosion, a temperature range between room temperature and up to $30'000$ K is covered.
In this range extremely complicated absorption spectra including all kinds of transitions occur,
and the radiation is far from equilibrium although the plasma can often be considered at LTE.
Last but not least, the geometries are usually of complicated three-dimensional
nature without much symmetry, as for instance in a electric circuit breaker.
More details are given by \cite{Jones1980}, \cite{Aubrecht1994}, \cite{Eby1998}, \cite{Godin2000},
\cite{Dixon2004}, and \cite{Nordborg2008}.\\
It is sufficient for our purpose to restrict the considerations to the radiation part for a given temperature profile,
for instance of a cylindrical electric arc in a gas in front of a plate with a slit (see Fig. \ref{arc2d}).
We may neglect scattering in the gas ($\sigma _{\nu}\equiv 0$) and mention that an electric arc consists of a very hot, emitting but
transparent core surrounded by a cold gas, which is opaque for some frequencies and transparent for others.
First, one has to determine the effective transport coefficients $\kappa^{\rm (eff)}_E$, $\kappa^{\rm(eff)}_F$, and $\chi (v)$,
with the above introduced entropy production minimization method. For simplicity,
we assume now that this is done and these functions are given simply by constant values listed in the caption of
Fig. \ref{arc2d}, and that $\chi (v)$ is well-approximated by Kershaw's VEF. Note that due to the low density in the hot arc core, the
effective absorption coefficient there is smaller than in the surrounding cold gas. Therefore, one expects that
the radiation in the arc center will exhibit stronger nonequilibrium than in the surrounding colder gas.\\
The energy density $E$ and the velocity vectors ${\bf v}={\bf F}/E $ obtained by a simulation with the commercial software
ANSYS$^{\textregistered}$ FLUENT$^{\textregistered}$
are shown in Figs. \ref{arc2d}. At the outer boundaries, homogeneous Neumann boundary conditions are used for
all quantities. The wall defining the slit is modelled as a material with
either a) high absorption coefficient or b) high scattering coefficient. The behavior of the velocity vector field clearly reflects
these different boundary properties. The $E$-surface plot shows the shadowing effect
of the wall when the arc radiation is focused through the slit. The energy densities $E$ along the $x$-axis are shown in
Fig. \ref{arcEandF} a) for the two cases. One observes the enhanced $E$ in the region of the slit for the scattering wall. The
energy flux in physical units, i.e., $cF$, on the screen in front of the slit is shown in Fig. \ref{arcEandF} b). The effect here is again
what one expects: an enhanced and less focused power flux due to the absence of absorption in the constricting wall.

\section{Summary and Conclusion}
\label{Conclusion}
After a short general overview on radiative heat transfer, this chapter has focused on truncated moment expansions of the 
RTE for radiation modelling.
One reason for a preference of a moment based description is the occurrence of the moments directly in the hydrodynamic equations for the
matter, and the equivalence of the type of hyperbolic partial differential equations for radiation and matter, which
allows to set numerical simulations on an equal footing.\\
The truncation of the moment expansion requires a closure prescription, which determines the unknown transport coefficients and
provides the nonequilibrium distribution as a function of the moments.
It was the main goal of this chapter to introduce the minimum entropy production rate closure, and to illustrate with the help of
the two-moment approximation that this closure is the one to be favored due to the following properties of the result:
\begin{itemize}
\item
It is exact near thermodynamic equilibrium, and particularly leads to the Rosseland mean absorption coefficients.
\item
It exhibits the required flux limiting behavior by yielding reasonable variable Eddington factors.
\item
It gives the expected results in the emission limit, and particularly leads to the Planck mean absorption coefficient.
\item
It can be generalized to an arbitrary number of and type moments.
\item
It can be generalized to particles with arbitrary type of energy-momentum dispersion (e.g. massive particles) and
statistics (Bosons and Fermions), as long as they are described by a linear BTE. In stellar physics, for instance,
neutrons or even neutrinos can be included in the analogous way.
\end{itemize}
The main requirement of applicability is that the particles be independent, i.e., they interact on the microscopic scale only
with a heat bath but not among each other. On a macroscopic scale, long-range interaction (e.g., Coulomb interaction)
via a mean field may be included on the hydrodynamic
level of the moment equations. Independency, i.e. linearity of underlying Boltzmann equation, has the effect that on the level of the
BTE (or RTE) nonequilibrium is always in the linear response regime. In this sense, all transport steady-states are 
near equilibrium even if $f_{\nu} $ strongly deviates from $f_{\nu}^{(eq)}$, and the entropy production rate optimization according to
\cite{Kohler1948} can be applied.


\end{document}